# Framework for additive manufacturing of porous Inconel 718 for electrochemical applications

Ahmad Zafari[a,1*], Kiran Kiran[b,1], Inmaculada Gimenez-Garcia[b], Kenong Xia[c], Ian Gibson[a], Davoud Jafari[a*]

[a]*Department of Design, Production and Management, Faculty of Engineering Technology, University of Twente, 7500AE, The Netherlands*

[b]*Department of Chemical Engineering and Chemistry, Eindhoven University of Technology, 5600MB, Eindhoven, The Netherlands*

*cDepartment of Mechanical Engineering, The University of Melbourne, VIC 3010, Australia*

***Corresponding authors:** ahmad.zafari@utwente.nl; davoud.jafari@utwente.nl

## Abstract

Porous electrodes were developed using laser powder bed fusion of Inconel 718 lattice structures and electrodeposition of a porous nickel catalytic layer. Laser energy densities of ~83–333 J/m were used to fabricate ~500 µm thick electrodes made of body centered cubic unit cells of 200–500 µm and strut thicknesses of 100–200 µm. Unit cells of 500 µm and strut thickness of 200 µm were identified as optimum. Despite small changes in feature sizes by the energy input, the porosity of >50% and pore size of ~100 µm did not change. In a subsequent step, we used nickel electrodeposition to create smaller scale pores on the electrode. The electrochemical performance of the electrodes for hydrogen/oxygen evolution reaction (HER/OER) was evaluated in a three-electrode setup. For HER, a much larger maximum current density of ~ −372 mA/cm$^2$ at a less negative potential of ~−0.4 V vs RHE (potential against reversible hydrogen electrode) was obtained in the nickel-coated samples, as compared to −240 mA/cm$^2$ at ~−0.6 V in the bare one, indicating superior performance of the coated sample. Conversely, OER exhibited minor performance differences upon application of the coating, indicating insignificant dependence of OER to surface composition and available surface.

[1]*These authors contributed equally to the work*

## 1. Introduction

Metallic porous materials play a crucial role in various industrial applications, including heat exchangers, structural components, absorbents, catalyst supports, and electrodes. The manipulation of their three-dimensional characteristics, such as porosity, pore size distribution, and thickness, is instrumental in defining their ultimate properties and functionality across different applications [1-5]. This significance is particularly notable for electrochemical technologies, where porous electrodes are responsible for numerous vital functions within the cell, influencing thermodynamics, kinetics, and durability. A prominent instance is the active research focused on water electrolysis, driven by the societal imperative of producing low-cost hydrogen without emissions. Within the electrolytic cell, the pivotal role of porous conductive materials, including porous transport layers and catalyst layers, becomes evident, distributing the liquid electrolyte, providing surfaces for electrochemical reactions, and incorporating open porous structures to facilitate the removal of generated gases [6, 7]. Consequently, the three-dimensional structure and surface properties of these materials emerge as decisive factors dictating the overall performance of the system. Traditionally, porous metallic materials are produced by powder/fiber sintering, and foaming [8-10]. However, these processes are still unable to precisely control pore size, morphology, and distribution, limiting the performance of the resultant systems due to inadequate electrolyte distribution or subpar electrolyte-electrode interface. Thus, there is a need for developing new material sets with specific micro/macro-structures and surface properties to enable the next-generation of electrochemical technologies.

Additive manufacturing (AM) has the potential to address the aforementioned challenges, since it can control the formation of pores to a great extent. Laser powder bed fusion (LPBF) is particularly of interest due to its relatively high spatial resolution in the range of 80-250 μm [11] depending on input parameters. Geometrically undefined porosity (GUP) in LPBF-fabricated structures is one of the most studied types of porosity engineered by manipulating laser parameters, namely laser type (pulsed and continuous waves), hatch distance, power, and scanning speed [12-14].

Generally, these parameters affect feature sizes (pores and wall thickness in here) by changing energy input and melt pool geometry. This has been demonstrated in various materials, including Inconel 718 [15], Inconel 625 [16], CL 20ES [17] and 17-4 PH stainless steel [18], leading to the porosity of ~10-40% in large specimens. Although such porous structures have multiple applications, such as breathable moulds [19] and heat pipes [20, 21], they are not suitable when high permeability and low ohmic resistance are required, as in the focus application of this study, including fuel cells or water electrolysers. For such applications, thin porous walls of 150-500 μm [22-24] are manufactured as a porous transport layer (PTL) or gas diffusion layer (GDL). Higher thicknesses normally lead to detrimental effects on permeability [25-27], sealing [27], and electrochemical performance due to higher electrical resistance and diffusion losses [26, 28, 29]. A recent study [30] showed that GUP of 17% with pore sizes of ~2-14 μm could be achieved in 1mm-thick 316 stainless steel walls by varying laser beam diameter, hatch distance and scanning strategy. However, GUP is still stochastic by nature, leading to random formation of pores with no absolute control over their locations and sizes. Further, a wall of 1 mm in thickness was still much thicker than the required thickness for PTL/GDL. Hence, further improvement in the LPBF process is essential to reduce the wall thickness below 500 μm.

Geometrically defined lattice structure porosity (GDLSP), commonly known as cellular/lattice structures, is another type of porosity widely produced using LPBF and electron beam powder bed fusion (EBPBF) [12, 14]. A unit cell repeats throughout a volume, producing a porous material in which the locations, sizes, and morphologies of the pores are initially determined by a digital model, making the porosity much more controllable than GUP. A variety of unit cells, such as body-centred-cubic (BCC), octahedron, and diamond, have been developed, with a majority investigated for biomedical and thermal management applications [31, 32]. However, cellular structures have been significantly less utilised in electrochemical systems, perhaps, due to their relatively large pores of > 100 μm – 1 mm [12, 14]. Though, recent studies, inspired by conventional porous GDLs (e.g., nickel foam/mesh, carbon paper) coated by suitable catalysts (e.g., nickel, platinium, paladium) for better electrochemical efficiency [33-37], have shown that a hybrid manufacturing consisting of AM and post-AM coating processes can close this gap. Several AM methods have thus far been used to produce GDLs, including LPBF of 316 stainless steel [38], EBLPF of Ti-6Al-4V [39], stereolithography (SLA) of polytetrafluoroethylene (PTFE) dispersion [40], and extrusion-based AM of various polymers, ceramics and metals [41-44]. In all these studies a secondary step of depositing such coatings as nickel, nickel-molibdinium, copper and carbon was added for different purposes. Although these works demonstrate an increasing interest in the use of AM in electrochemical systems, there are still some shortcomings. For example, the pore sizes were of the order of several milimiters, and the thicknesses and strut sizes of those fabricated by EBLPF and materials extrusion were quite high (i.e., several milimeters), giving rise to high ohmic resistance and sealing problems. Nonetheless, LPBF is more promising because of its high spatial resolution and controllability through digital models and input process parameters. Therefore, it is crucial to conduct further research on cellular structures to thoroughly investigate the capabilities of the current LPBF technology, while also identifying its limitations and determining the optimal combination of designs, pore sizes, and laser parameters.

The primary objective of this study is to regulate pore characteristics, specifically targeting the optimisation of surface area for facilitating electrochemical reactions in alkaline media relevant to water electrolysis. Departing from conventional foaming methods, we utilised LPBF to craft lattice-structured electrodes. A digital model was employed for the preprograming of pore geometry, electrode thickness, and pore size and distribution. The design criteria were centered on attaining the finest pores and the highest porosity within a thin electrode. Consequently, electrodes with dimensions of approximately 500 µm, featuring uniform pore distribution at around 100 µm and exceeding 50% porosity, were produced through LPBF of metals. This contrasts with prior AM-fabircated electrodes characterised by thicker dimensions on the order of 1 mm and much larger pores. For achieving even finer pores at a micron and submicron scale, an additional layer of porous nickel catalyst was introduced, resulting in a gradient of pores within the electrodes. In718, a nickel-based superalloy initially developed for high-temperature applications due to its commendable strength, creep resistance, and fatigue life, was chosen for this study. Although previous research on additive manufacturing (AM) of In718 primarily focused on mechanical properties [45-48], its chemical composition closely aligns with commercial nickel foams and is compatible with effective catalysts such as nickel and nickel-iron [49, 50]. Moreover, In718 exhibits excellent printability, and its commercially available powder renders it advantageous for cost-effective fabrication of high-quality electrodes through LPBF. To increase the surface area for electrochemical reactions, a nickel layer was further deposited using dynamic hydrogen bubble templating. Various designs and laser parameters were employed to explore LPBF limitations and establish an optimal processing procedure. The electrochemical performance of the electrodes was assessed by measuring polarisation measurements in a 3-electrode setup for both the hydrogen evolution reaction (HER) and oxygen evolution reaction (OER). This study not only lays the groundwork for the application of LPBF in the energy industry but also provides guidelines for advancing the technology until its full potential is realised.

## 2. Materials and experimental procedures

Experimental procedures in the present work include two major parts: sample fabrication (section 2.1) divided into LPBF (2.1.1) and electrochemical deposition (2.1.2), and characterisation (section 2.2). The methods involved in the latter are described in sections 2.2.1 to 2.2.3 detailing optical and electron microscopy, pore size and porosity measurements, and electrochemical testing, respectively.

### *2.1. Porous materials fabrication and functionalisation*

#### *2.1.1. Laser powder bed fusion (LPBF)*

Gas atomised In718 powder (15-45 µm) supplied by Renishaw was used for manufacturing porous and solid structures using the Renishaw AM250 LPBF system with a laser diameter of 66 µm. The 3D printing was performed in a reduced build volume (RBV) attachment using a stripe scanning strategy (stripe size of 5 mm) and rotation angle of 67° between successive layers of 60 µm in thickness. No preheating was applied on the 304 stainless steel build plate, and the printing chamber was under an ultrahigh purity Ar gas atmosphere with an oxygen content set at a maximum of 1000 ppm.

To obtain high porosity, a BCC unit cell was used and digital files were generated in Netfabb, as seen in Fig. 1. Thin walls of 10 mm × 10 mm × 0.5 and 1 mm (width × height × thickness) were produced for design and microstructural analyses, and a set of 15 mm × 17 mm × 0.5 mm samples were also fabricated for HER and OER examinations. Design parameters included the strut diameters of 0.1 and 0.2 mm, unit cell sizes of 0.2, 0.3 and 0.5 mm, and wall (electrode) thicknesses of 0.4–1 mm, as summarised in Table 1. Specimens were designated as Sx-y-z, where S stands for specimen, and x, y and z denote the values of strut diameter, unit cell size, and electrode thickness in millimetre, respectively. Specimens of 0.5 mm thick were made of one layer of unit cells (middle,

Fig. 1) and the rest contained two layers (right, Fig. 1). The electrodes were printed on top of solid bases as shown in Fig. 1 for easier removal from the build plate.

**Table 1 Design parameters using BCC unit cells for manufacturing porous walls. Single and double layers refer to the number of unit cells across the walls, as shown in Fig. 1.**

| Specimen ID | Strut size (mm) | Unit cell size (mm) | Wall thickness (mm) |
|---|---|---|---|
| S0.1-0.5-0.5 | 0.1 | 0.5 | 0.5 (single layer) |
| S0.1-0.5-1.0 | 0.1 | 0.5 | 1.0 (double layer) |
| S0.2-0.5-0.5 | 0.2 | 0.5 | 0.5 (single layer) |
| S0.1-0.3-0.6 | 0.1 | 0.3 | 0.6 (double layer) |
| S0.1-0.2-0.4 | 0.1 | 0.2 | 0.4 (double layer) |

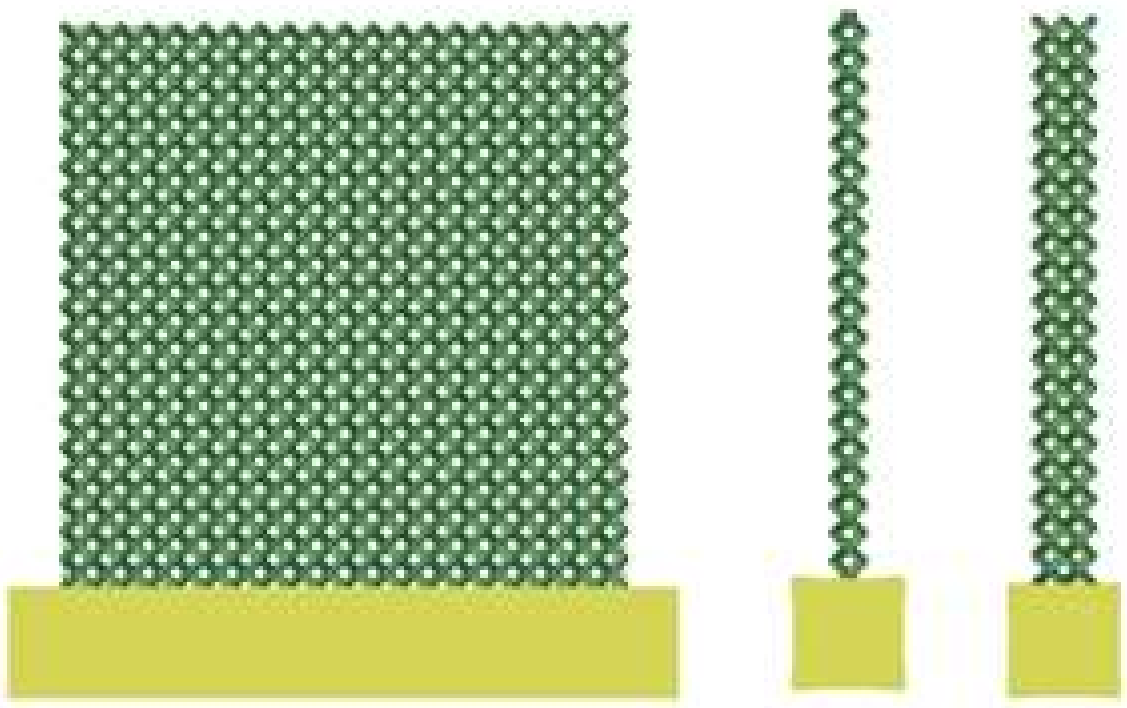

**Fig. 1 Digital illustration of a porous wall with BCC unit cells, showing the front view (left), one single layer (middle) and two layers (right) of the unit cells across the wall thickness. The walls were placed on solid bases to be printed together.**

To manipulate feature sizes, a wide range of powers and scanning speeds (by changing exposure time (ET) and point distance (PD)) were used, as listed in Table 2, resulting in volumetric energy densities (VED) of ~10 to ~40 J/mm$^3$ estimated using

$$VED = \frac{P}{v \cdot h \cdot l_t} \tag{1}$$

and linear energies ($E_l$) of ~83 to 333 J/m calculated by

$$E_l = \frac{P}{v} \tag{2}$$

where $P$ denotes power, $v$ scanning speed (= PD / ET), $h$ hatch distance, and $l_t$ layer thickness.

The same laser parameters were also used to produce a set of solid truncated cones with a height of 10 mm, and smaller and larger circles of 5 and 10 mm in diameter, respectively.

**Table 2 Laser parameters used to produce porous electrodes and truncated cones. Laser parameters are in three categories of varied power, exposure time and point distance, with the last two determining scanning speed, while the remaining parameters are constant.**

| Category | $h$ (µm) | PD (µm) | ET (µs) | $v^{(1)}$ (mm/s) | $P$ (W) | VED (J/mm$^3$) | $E_l$ (J/m) |
|---|---|---|---|---|---|---|---|
| | 140 | 60 | 100 | 600 | 200 | 39.7 | 333.3 |
| $P$ = 70-200 W$^{(2)}$ | 140 | 60 | 100 | 600 | 160 | 31.8 | 266.7 |
| | 140 | 60 | 100 | 600 | 140 | 27.8 | 233.3 |
| | 140 | 60 | 100 | 600 | 120 | 23.8 | 200 |
| | 140 | 60 | 100 | 600 | 100 | 19.8 | 166.7 |
| | 140 | 60 | 100 | 600 | 80 | 15.9 | 133.3 |
| | 140 | 60 | 100 | 600 | 70 | 13.9 | 116.7 |
| ET = 40-90 µs | 140 | 60 | 90 | 666.67 | 200 | 35.7 | 300 |
| | 140 | 60 | 80 | 750 | 200 | 31.8 | 266.7 |
| | 140 | 60 | 70 | 875.14 | 200 | 27.8 | 233.3 |
| | 140 | 60 | 60 | 1000 | 200 | 23.8 | 200 |
| | 140 | 60 | 50 | 1200 | 200 | 19.8 | 166.7 |
| | 140 | 60 | 40 | 1500 | 200 | 15.9 | 133.3 |
| $P$ = 100 W; ET = 50-90 µs$^{(3)}$ | 140 | 60 | 80 | 750 | 100 | 15.9 | 133.3 |
| | 140 | 60 | 70 | 875.14 | 100 | 13.9 | 116.7 |
| | 140 | 60 | 60 | 1000 | 100 | 11.9 | 100 |
| | 140 | 60 | 50 | 1200 | 100 | 9.9 | 83.3 |
| PD = 70-90 µm$^{(4)}$ | 140 | 70 | 100 | 700 | 200 | 34.0 | 285.7 |
| | 140 | 80 | 100 | 800 | 200 | 29.8 | 250 |
| | 140 | 90 | 100 | 900 | 200 | 26.5 | 222.2 |

(1) Scanning speed = point distance /exposure time.
(2) Parameters in the first row were provided by Renishaw.
(3) These parameters were only used to fabricate a set of S0.1-0.3-0.6 specimens.
(4) These parameters were only used to fabricate a set of cones.

### *2.1.2. Electrochemical deposition*

An additional nickel-based catalyst layer was deposited on selected LPBF-fabricated electrodes. Ammonium chloride, $NH_4Cl$ (1.5 M, Fluka, > 99%) served as a supporting electrolyte, and nickel (II) chloride hexahydrate, $NiCl_2.6H_2O$ (0.1 M, sigma-aldrich, ReagentPlus®) as the metal precursors for the plating step. All chemicals were used without further purification. Milli-Q water (Millipore, 18.2 MΩ cm) was used for the preparation of all solutions. A two electrode setup was employed for electrodeposition (Table S1). LPBF-fabricated cellular structures served as working electrodes, a platinium (Pt) mesh as the counter electrode, and a Delta Elektronika ES030-5 (30V, 5A) was utilised as a current source. To perform the deposition on a well-defined and reproducible geometric surface area of 1 $cm^2$, the LPBF fabricated electrode was masked with an insulating PTFE tape. Herein, the dynamic hydrogen bubble template (DHBT) assisted method was employed for preparation of catalyst, as described elsewhere [51-53]. A chloride bath was used to form a porous catalyst layer adapted from Guo et al. [54]. The porous nickel layer was deposited at cathodic current density of −1 A $cm^{-2}$ for 100 s.

## *2.2. Physicochemical characterisation*

### *2.2.1. Optical and electron microscopy*

Optical microscopy (OM; Keyence VHX 7000) was performed to measure strut diameters. Porous electrodes were first hot mounted in black bakelite resin and sectioned by SiC sandpapers of grit sizes 1000 and 2000 followed by polishing using 1 μm diamond and active oxide polishing solutions. This process was repeated three times and an average of the measurements obtained from the three sections was reported as the strut diameter.

Melt pool boundaries were revealed by immersing polished specimens in an etchant containing 37% HCl and $H_2O_2$ at room temperature for 8-10 seconds. OM images were then used for melt pool size measurements. The widest and deepest dimensions of a melt pool were recorded as the melt pool width and depth.

Scanning electron microscopy (SEM; FEI Teneo and Quanta FEG) was employed for observations of as-LPBF specimens and nickel layers on the coated electrodes. Energy-dispersive X-ray spectroscopy (EDS) was used to map the elemental nickel distribution on the coated electrodes. To assess the crystallographic structure of the coated electrodes, X-ray diffraction (XRD) was carried out using a tabletop Bruker D8 advanced diffractometer. For the XRD analysis, a Cu Kα radiation source (λ= 0.1540 nm, 40 mA) operated at 40 keV was used. XRD spectra were recorded in reflection mode in steps of 0.02° $sec^{-1}$ with 2θ values ranging from 10 to 100°. To exclude nickel diffraction peaks originating from the support, the nickel coating was deposited onto a graphite substrate (99.8%, Alfa Aesar, 0.13 mm thickness). The obtained XRD patterns were analysed and compared to COD (crystallography open database).

*2.2.2. Density and porosity measurements*

The densities of the cones were measured using Archimedes' principle and the corresponding relative densities were calculated by normalising them against the bulk density of In718, which is 8.19 $g/cm^3$ [55].

The porosity and density of porous structures were measured using [30]

$$Porosity = 1 - \frac{\rho_{por}}{\rho_b} \quad (3)$$

and

$$\rho_{por} = \frac{m_{por}}{A_{por} \cdot t_{por}} \quad (4)$$

where $\rho_{por}$ is the density of porous material, $\rho_b$ is nominal density of a bulk In718 (8.19 g/cm3 [55]), and $m_{por}$, $A_{por}$ and $t_{por}$ are mass, area and thickness of the porous material, respectively. $m_{por}$ was measured using a high accuracy balance (Mettler Toledo), $A_{por}$ was calculated by the image processing toolbox in MATLAB from OM images of the samples weighed, and $t_{por}$ by a micrometre.

*2.2.3. Electrochemical testing*

Electrochemical analysis was conducted on as-LPBF cellular structures before and after depositing nickel, as an additional catalytic layer, using a Biologic potentiostat (VSP-3e). As a proxy for the relevant electrochemical reactions in alkaline water electrolysers, we here perform 3-electrode polarisation testing of both HER and OER. Potassium hydroxide, KOH (1M, Sigma-Aldrich), was used as the electrolyte. The as-printed In718 and nickel-coated In718 were employed as working electrodes (WE), reversible hydrogen electrode (RHE, HydroFlex) served as the reference electrode (RE) and Pt mesh was the counter electrode (CE). It should be noted that although the electrochemical dissolution-deposition of Pt might affect HER performance [56, 57], it has been shown that this effect can only be significant in acidic media [58], while our system is highly alkaline, and therefore, unaffected by this particular concern. Further, graphite was not selected as CE due to its instability and the risk of the production of CO and $CO_2$, which would affect the iV measurements [59, 60]. Prior to the electrochemical experiment, the solution was deaerated with argon for ~5-10 min and then an argon blanket was kept over the solution during measurements. In order to check the activity of catalysts for HER and OER in a classical three electrode cell, the following protocol was followed. The working electrodes were subjected to 100 cycles of cyclic voltammetry (CV) in the potential range of −0.3 to 1.6 V vs RHE as a pretreatment step. This step was performed to ensure the testing conditions for all the experiments were similar. Electrochemical impedance spectroscopy (EIS) measurements were performed by applying constant potential by varying the frequency from 300

KHz to 100 mHz, with 8 points per decade. EIS measurements were done before the electrochemical reaction (HER and OER) to quantify the ohmic resistance, identified as the high-frequency intercept. Afterwards, cyclic voltammetry was performed at a scan rate of 50 mV/s in the potential range of −1 V to 2.3 V vs RHE. To obtain the polarisation curves, the electrodes were first subjected to a potential range in HER region (−0.2 to −1 V vs RHE) with a step potential of 50 mV. Each potential step was held for 1 min to reach a steady state and the corresponding current response was recorded. The obtained values of the current for each potential were averaged for the last 50% of the recorded data points, and then normalised by the geometric surface area (1 $cm^2$) to obtain the values of current density ($j$). The measured potential was corrected with iR compensation obtained from the ohmic resistance of the electrochemical cell. The corrected potentials were then plotted against the corresponding current densities to produce polarisation curves. The same procedure was repeated for OER, in a potential range between 1.5 V and 2.3 V vs RHE.

## 3. Results and discussion

This section presents the main findings of the study obtained from different electrode designs produced using a wide range of laser parameters leading to an optimum manufacturing procedure. The electrochemical behaviours of the optimum design are also discussed as a proof-of-concept. Particularly, laser parameters are first introduced in 3.1, followed by design optimisation of cellular structures in 3.2 and the effects of laser parameters on cells in 3.3. Finally, electrochemical properties are discussed in 3.4.

### *3.1. Optimisation of laser parameters for bulk*

A wide range of laser parameters inducing various energy inputs were first used to produce bulk structures. This would eventually provide necessary information about the right parameters for LPBF of cellular structures. This would also be fundamentally significant to understand differences between additively manufactured bulk and porous materials.

Fig. 2 shows the relative densities of the LPBF-fabricated truncated cones as a function of VED. The highest destiny of > 99% was obtained at ~40 J/mm$^3$, and it reduced to ~ 80% at ~14 J/mm$^3$. The reduction of the density was attributed to an increase in LPBF defects, particularly lack of fusion and large unmelted regions at lower VEDs, as typified in Figs. 3a-b. As seen in Fig. 3a, a VED as low as 16 J/mm$^3$ resulted in large unmelted regions, as well as partially melted/sintered particles which were removed during grinding and polishing. By doubling the VED, the unmelted/sintered regions were significantly reduced and better fusion between layers and particles occurred (Fig. 3b). Lack-of-fusion sites were eventually eliminated by increasing VED to ~40 J/mm$^3$ in which only small isolated pores existed (not shown here).

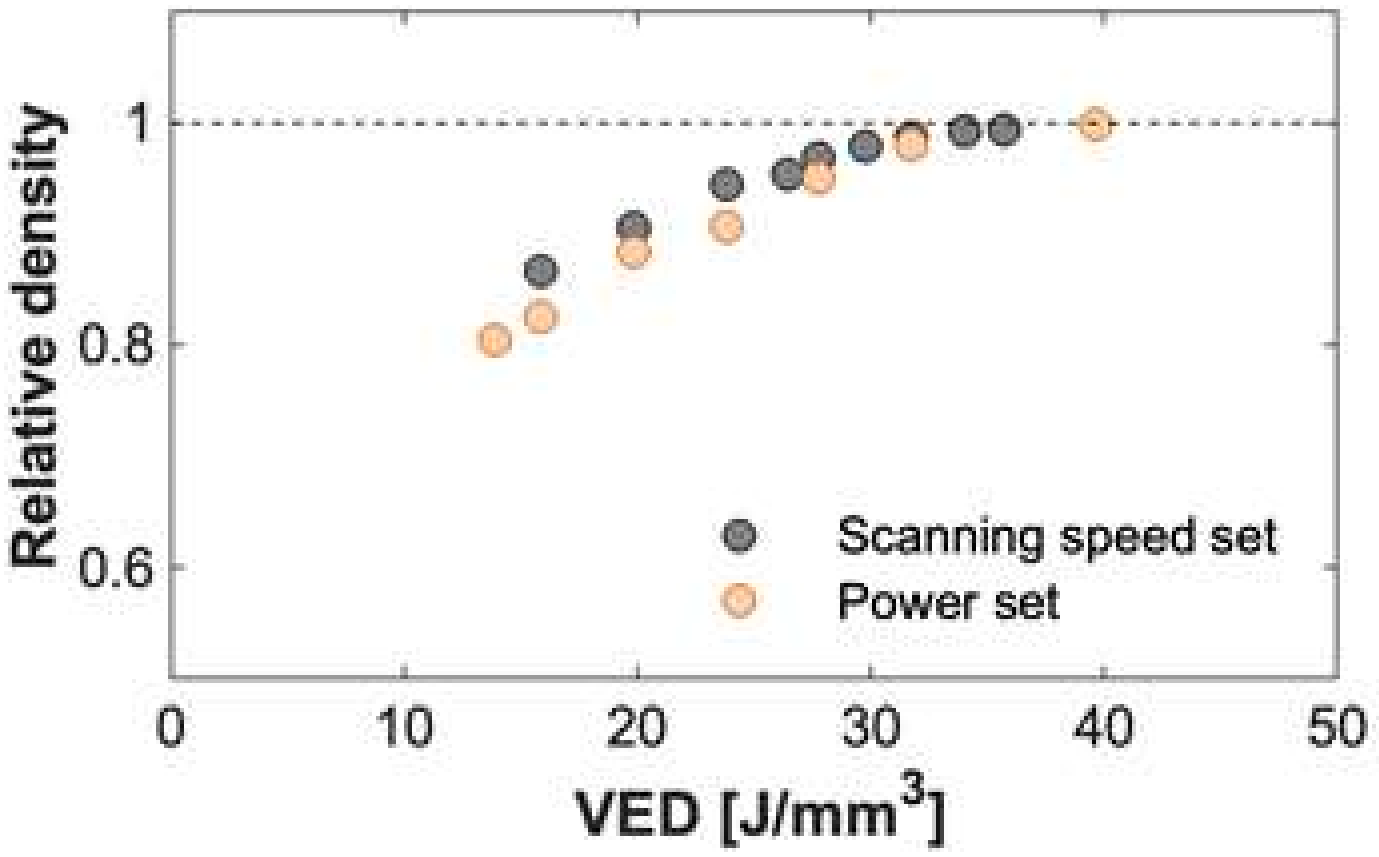

**Fig. 2 Relative density as a function of VED varied by scanning speed (*v* of 600-1500 mm/s and the constant *P* of 200 W) and power (*P* of 70-200 W and the constanct *v* of 600 mm/s), as indicated by two different colours in the legend, showing the highest density of > 99% at 40 J/mm$^3$ and a quick drop by decreasing VED below 35 J/mm$^3$ down to ~80% at ~14 J/mm$^3$. The dashed line represents a fully dense material with a relative density of 1.**

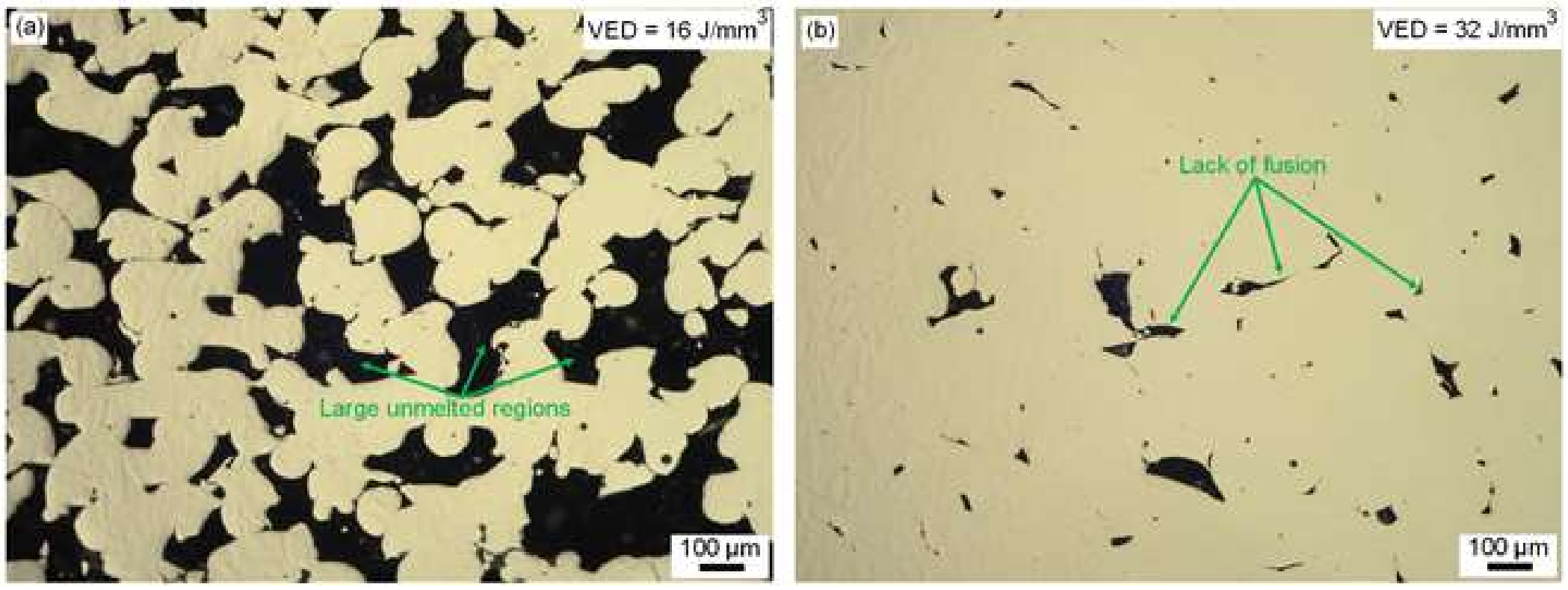

**Fig. 3 OM of the as-polished cones printed using *P* of (a) 80 W (VED = ~16 J/mm$^3$) and (b) 160 W (VED = ~32 J/mm$^3$), showing a substantial change in the quantity and sizes of defects from significantly large unmelted/sintered regions in (a) to much smaller lack fusion sites in (b).**

*3.2. Design optimisation*

Before discussing the effect of laser parameters on porous structures, an optimum design should first be identified. For this purpose, the laser parameters leading to the densest structure at the highest VED of 40 J/mm$^3$ (Fig. 2 and Table 2) were used for the first few attempts. S-0.1-0.5-0.5 was the first printed part, but it turned out to be an unsuccessful trial as the part was completely destroyed during printing. The failure was attributed to the inability of the small struts to withstand thermal stresses and forces applied by the powder recoater pushing powder to them. Hence, it was necessary to create more supporting structures to prevent the fracture of the struts. Several approaches were taken: (i) to add a second layer of unit cells, so that each unit cell supports the neighbouring one (S0.1-0.5-1.0); (ii) to reduce the unit cell size in order to reduce the length of the struts, and therefore, the magnitude of torques applied on them by the recoater (S0.1-0.2-0.4 and S0.1-0.3-0.6); and (iii) to double the diameter of the struts (S0.2-0.5-0.5). S0.1-0.5-1.0 was only partially printed and fractured to pieces during powder recovery and its separation from the build plate, indicating that the second layer of the unit cells of 500 µm was still inadequate to maintain the structural integrity of the wall during printing. Figs. 4a-e show the resulting specimens obtained from the implementation of the second and third approaches above. S0.1-0.2-0.4 was completely printed (Fig. 4a), but it was considered unsuccessful, because the structure was nonuniform, and more importantly, no light was passing through it, indicating that the wall was solid, instead of porous. On the contrary, a slightly larger unit cell size in S0.1-0.3-0.6 led to a porous structure (Fig. 4b-c). However, Fig. 4c, a closeup of the framed area in Fig. 4b, shows that pores were nonuniform and unit cells were substantially distorted beyond recognition. The best printability was achieved by increasing the strut diameter in S0.2-0.5-0.5, which appeared to contain periodic and reasonably uniform pores, and the distortion of the unit cells was insignificant (Figs. 4d-e).

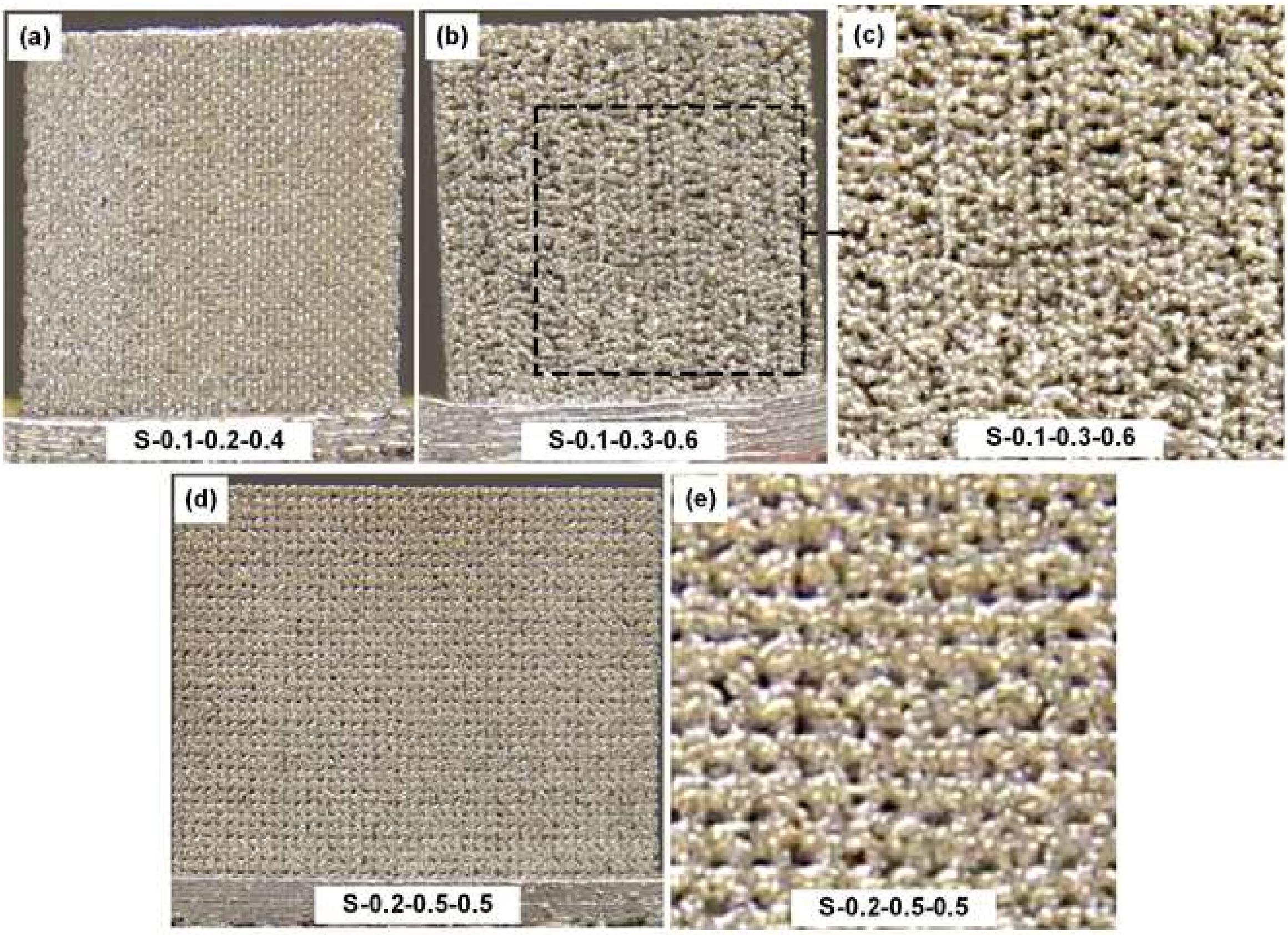

**Fig. 4 Photographs of (a) S-0.1-0.2-0.4 (double layer), (b) S-0.1-0.3-0.6 (double layer), (c) a closeup of the framed area in (b), (d) S-0.2-0.5-0.5 (single layer), and (e) a closeup of S-0.2-0.5-0.5, showing nonuniformity of the products in (b-d), and optimum printing in (d) and (e) with periodic geometrically defined pores.**

More detailed observations were made using SEM, as shown in Fig. 5. S0.1-0.2-0.4 in Fig. 5a-b was clearly a pore-free solid wall. The nominal pores in this design were even smaller than the melt pools of > ~100-200 μm (will be detailed later). As a consequence, the molten material filled the designed pores, leading to the formation of a solid wall. Similar to the macrographs in Figs. 4b-c, a porous structure was seen in S0.1-0.3-0.6 (Fig. 5c), but completely nonuniform and dissimilar to the digital design. This was a result of unit cells collapsing on top of each other and complete closure of pores in some areas, as typified in Fig. 5d. S0.2-0.5-0.5 in Fig. 5e, on the other hand, contained uniformly distributed pores, and X-shaped unit cells (solid lines).

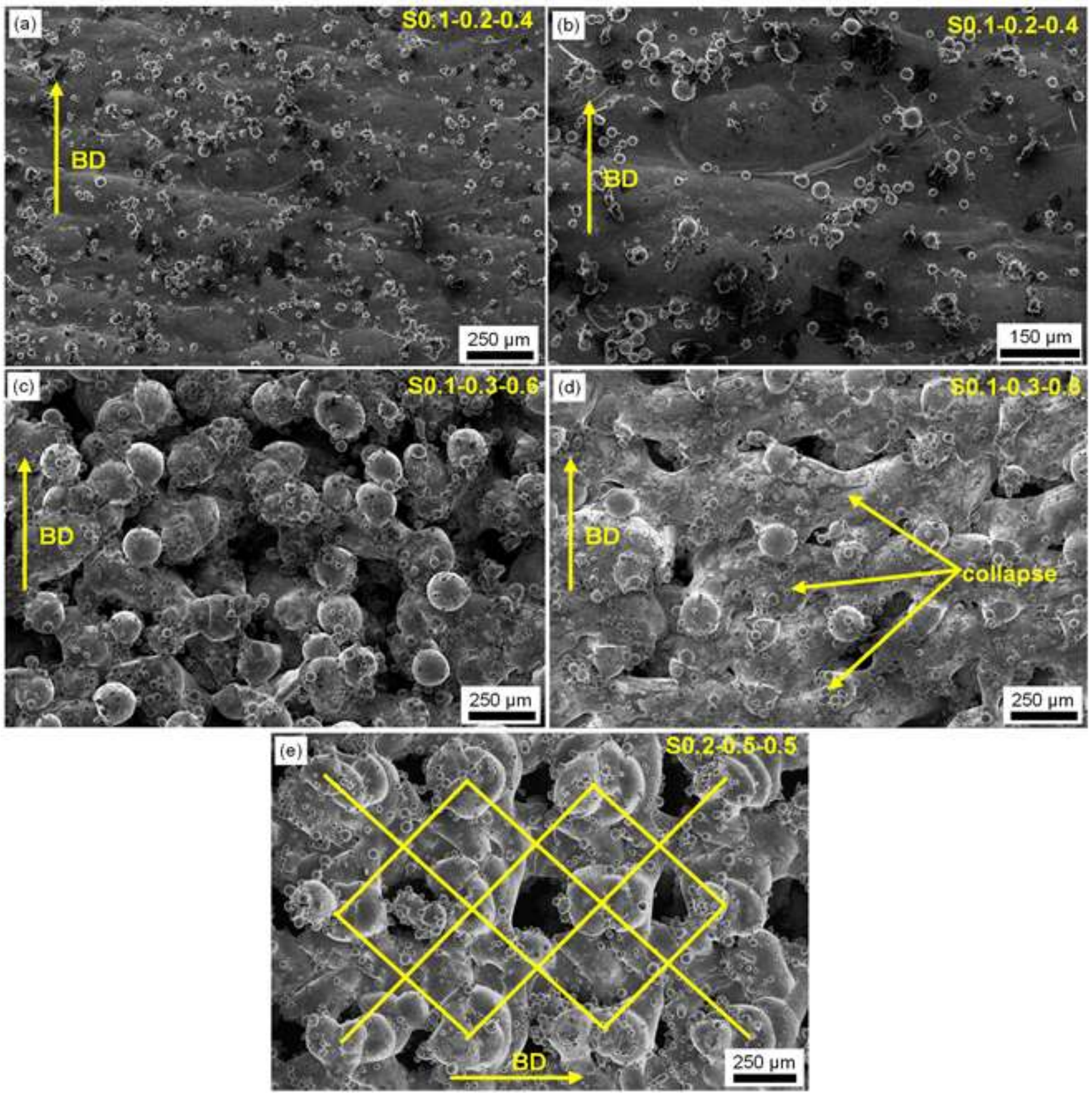

**Fig. 5 SEM micrographs of (a, b) S0.1-0.2-0.4 at two different magnifications, showing a solid wall free of pores, (c) S0.1-0.3-0.6 containing pores of the order of 100 μm, but distributed nonuniformly, (d) struts and nodes (intersections of struts) collapsed on top of each other in S0.1-0.3-0.6, giving rise to their coalescence, and (e) S0.2-0.5-0.5 with uniformly distributed, periodic pores, as well as clearly observable X-shaped unit cells (illustrated using solid lines). BD stands for building direction.**

Cross sections of the unit cells of S0.1-0.3-0.6 in Figs. 6a-c revealed the root cause of the collapses observed in Fig. 5d. Although a few struts and unit cells were identified (solid lines in Fig. 6a), the majority of them were melted together to create much larger melt zones, which eventually solidified into large solid areas (selectively circled in Figs. 6a-b). Fig. 6c shows a typical representative microstructure of such regions dominated by large dendrite arms with no visible melt pool boundaries. One explanation is that the struts of 100 µm in a small unit cell of 300 µm were too close to each other, leaving a small vacancy in the unit cell. As a consequence, the heat accumulation induced by printing 8 struts per unit cell would be high enough to melt the powder inside the vacancy. The heat accumulation would be even more significant when two rows of unit cells were printed side by side, as was the case in S0.1-0.3-0.6. When this extra melt zone in the vacancy coalesced with the melt pools in the struts, a much larger melt with a shape different from the original melt pools would be produced. Such large melt zones would be heavy and collapse on the previously solidified layers towards the gravity direction, leading to the substantial distortion shown in Figs. 4b-c and 5c-d. A similar phenomenon could also occur during LPBF of S0.1-0.2-0.4. The only difference was that a smaller unit cell contained a much smaller vacancy, and thus, a large molten zone would completely fill any pores. Such excessive melting was not prevented even by reducing VED to 10-16 J/mm$^3$ ($E_l$ of 83-133 J/m; Figs. 6d-e), indicating that the unit cells smaller than 500 µm were not printable by LPBF. These results led to the conclusion that S0.2-0.5-0.5 with periodic pores and recognisable X-shaped unit cells was the optimum design.

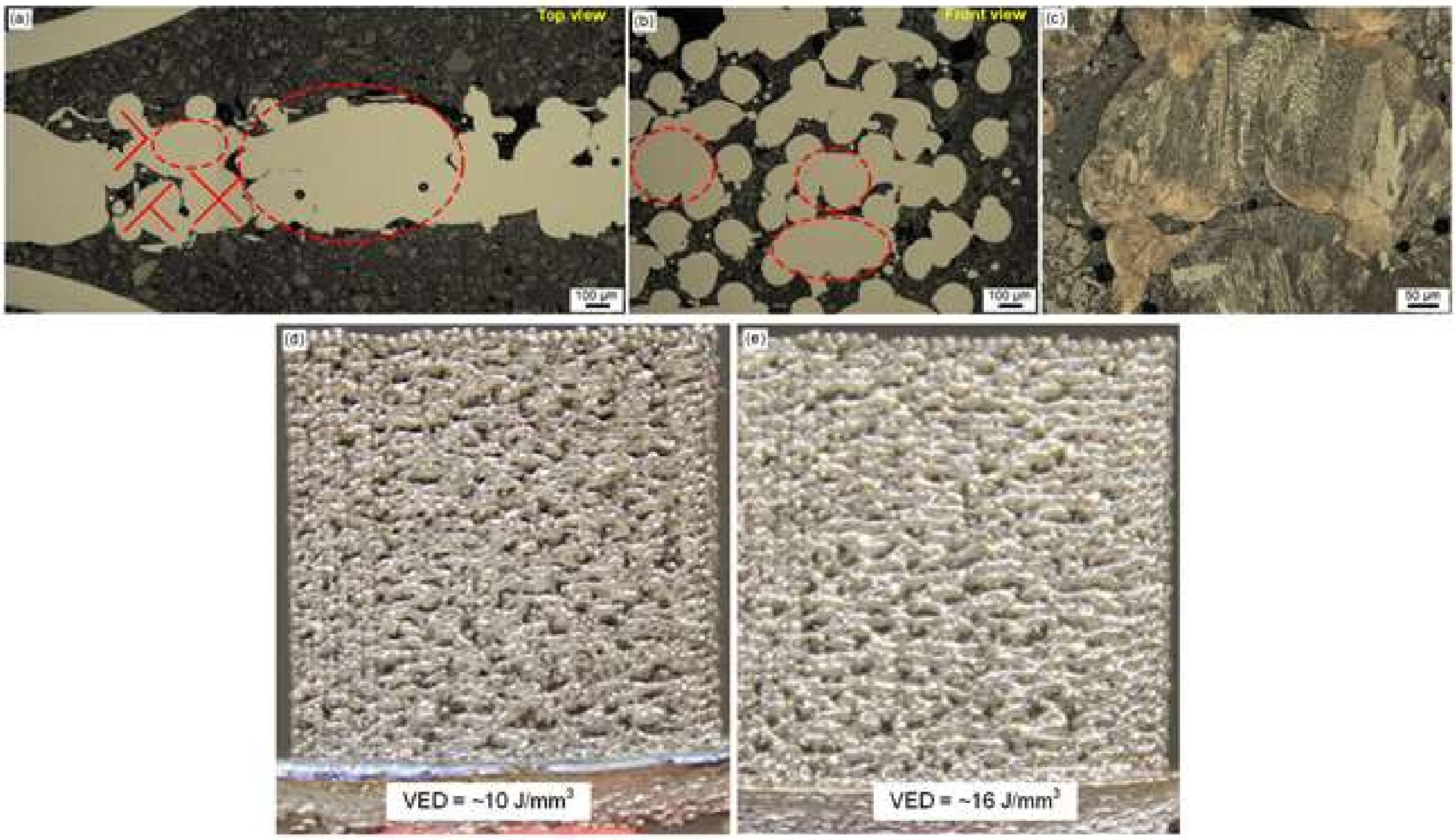

**Fig. 6 (a-c) OM images of the cross sections of S0.1-0.3-0.6, showing (a) top view and (b) front view with some struts recognisable (solid lines) while a majority of the unit cell coalesced to large solid regions (selectively circled), and (c) microstructure of the coalesced unit cells containing dendrites without any visible melt pool boundaries. (d) and (e) photographs of S0.1-0.3-0.6 fabricated using a low power of 100 W and low exposure times of 50 and 80 µs, leading to VEDs of ~10 J/mm$^3$ ($E_l$ = 83 J/m) and 16 J/mm$^3$ ($E_l$ = 133 J/m), respectively. The same unit cell coalescence and collapse are shown in (d-e) as those observed in Fig. 5c-d.**

*3.3. Effect of laser parameters on S0.2-0.5-0.5*

*3.3.1. Strut diameter and melt pool analysis*

S0.2-0.5-0.5 was printed using different laser powers and scanning speeds (Table 2). The strut diameters obtained from each set of parameters were measured using OM image analysis. As explained earlier, the analysis was conducted on three sections, and an average of the measurements was calculated as the strut diameter. Fig. 7 shows a representative image of one of the sections, revealing struts and nodes (intersections of struts).

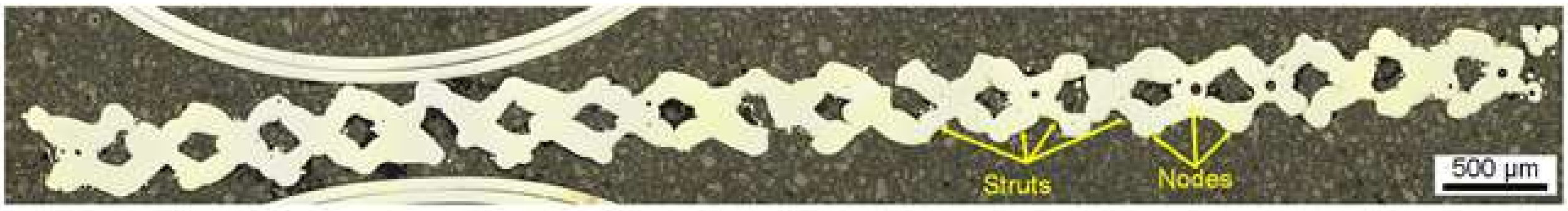

**Fig. 7 Typical representative OM image of a section of S0.2-0.5-0.5 fabricated using *P* = 100 W.**

Figs. 8a-c show the variations of strut diameters with increasing power, scanning speed (i.e., reducing exposure time), VED and $E_l$, respectively. It is worth noting that the strut diameters in Fig. 8 can be lower than their actual values because the struts might not have cut through their thickest sections. Also, the struts were not in one plane and there was an angle between the actual plane containing the struts and the plane of the view. Nonetheless, the data in Fig. 8 show a clear ascending trend of strut diameters with increasing energy input induced by either a higher power or slower scanning speed. A higher VED or $E_l$ leads to larger melt pools, and thus, larger features in the cellular structures. This was also experimentally proven by measuring depths and widths of the melt pools at different powers, scanning speeds, VED and $E_l$, as shown in Figs. 9a-c, indicating larger melt pools

for higher energy inputs. This is consistent with a previous study [61] that showed melt pool width ($b$) and strut diameter ($d$) were directly proportional to the square-root of $E_l$ (i.e., $d \propto b \propto \sqrt{E_l = P/v}$). However, the increasing trend in strut diameters was not substantial (i.e., from 100±25 to 158±34 μm), despite a significant increase in $E_l$. Although it is still unclear that why the strut diameters were only slightly increased, this could be ascribed to two possible scenarios. First, struts were small and contained a small amount of melt in each layer, leading to quick solidification, thus preventing the expansion of the melt to its full potential size. Second, as stated above, strut diameter ($d$) is a function of $\sqrt{E_l}$ [61], and as a result, the increasing rate of $d$ ($\dot{d} = \frac{1}{2\sqrt{E_l}}$) reduces with increasing $E_l$. In other words, above a certain range of $E_l$, $d$ and $b$ would not change significantly, and to observe big differences even lower energy should have been applied, which was outside the capacity of the LPBF machine used.

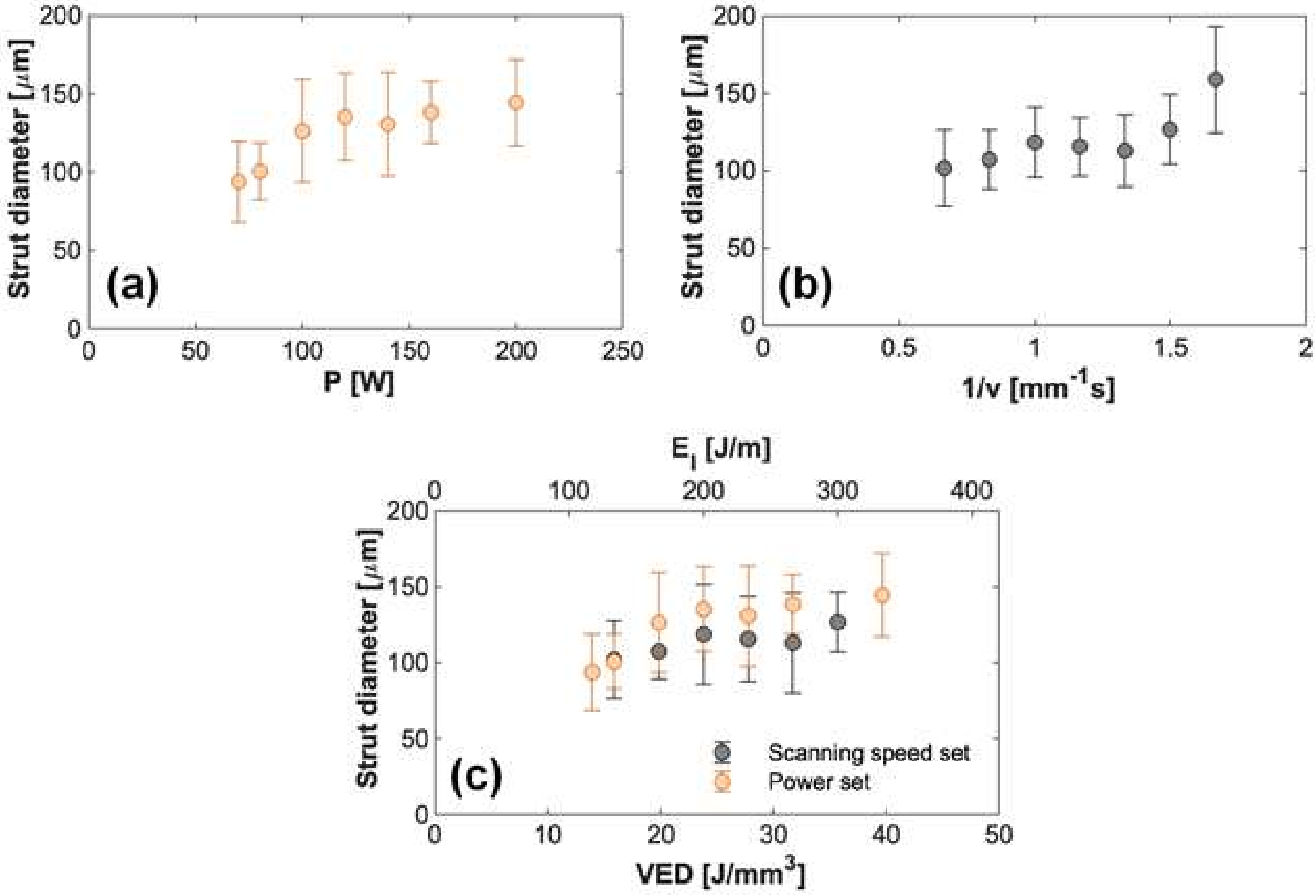

**Fig. 8 Variation of strut diameter with increasing (a) *P* (power), (b) 1/*v* (1/scanning speed), and (c) VED and *El* for different sets of power and scanning speed, showing an increase in the diameter with energy input stemmed from either a higher power or lower scanning speed.**

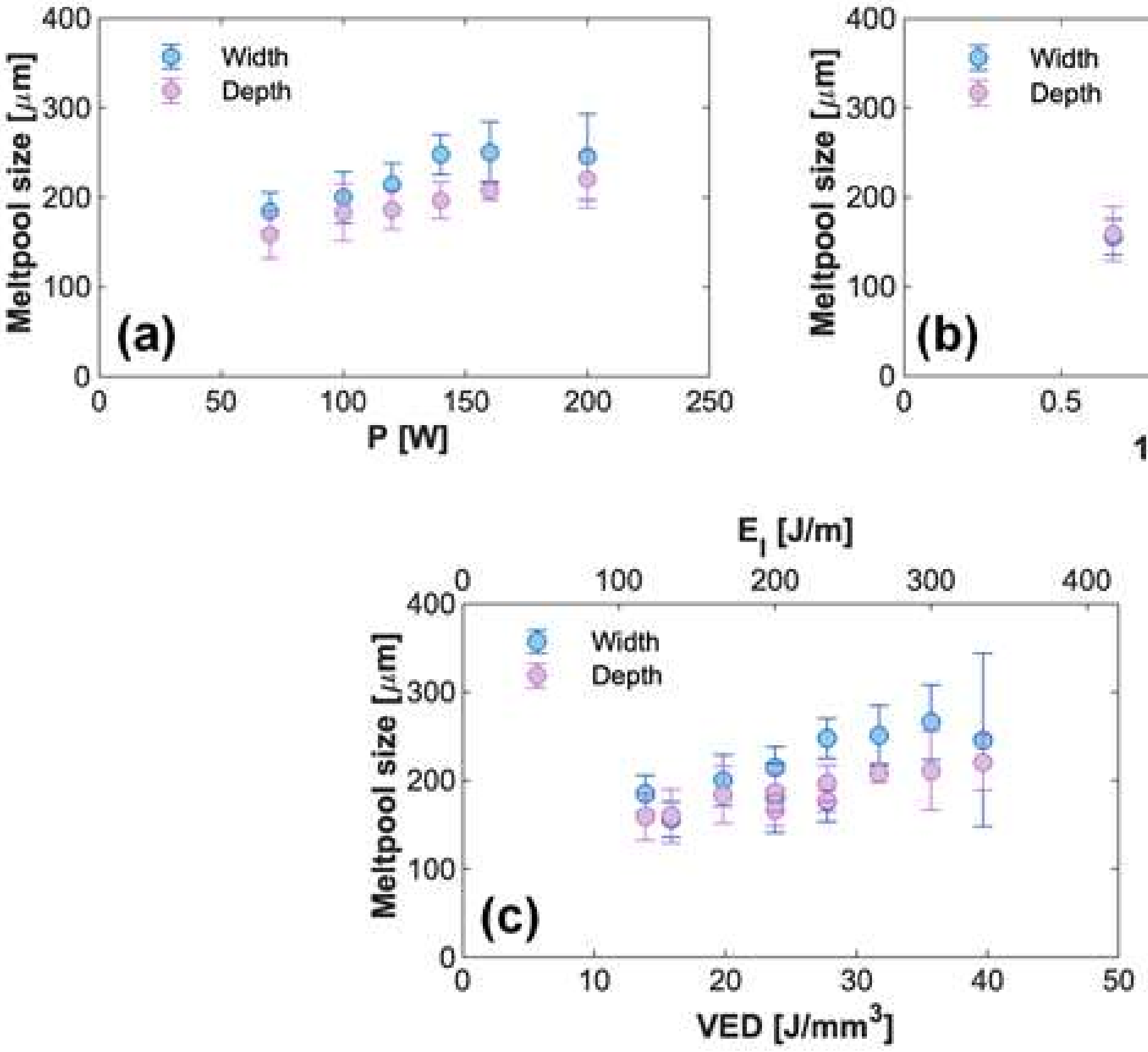

**Fig. 9 Melt pool size (width and depth) with respect to (a) *P* (power), (b) 1/*v* (1/scanning speed), and (c) VED and *El*, showing larger depths and widths at higher values of all the three parameters.**

On a side note, low VEDs did not lead to lack of fusion and sintering in cellular structures, which was in contrast to what was initially expected, since they occurred in bulk structures (Fig. 3). Figs. 10a-c show melt pools (selectively delineated by dashed lines) in a cone (i.e., bulk) and S0.2-0.5-0.5 (i.e., porous structure) fabricated using *P* of 140 and 70 W, respectively. Fig. 10a clearly shows shallower melt pools in the bulk (i.e., depth of ~109.8±23, and width of ~227.5±37) than those in the cellular structure (Figs. 10b-c and Fig. 9a), despite LPBF of the former at a much higher *P*. In fact, some of the melt pools in Figs. 10b-c were deep enough to create keyhole pores. This could be attributed to the potentially different conduction rates in bulk and cell structures. In cones, a much larger solid area was available for the heat to sink, which in turn increased the heat dissipation rate. Indeed, it was shown that thermal conductivity, which determined the rate of heat transfer, was directly proportional to the volume fraction of solid in cellular structures (i.e., $V_f$ = the ratio of solid material to free space) [62]. If a bulk is assumed to be a cell with almost no free space, $V_f$ approaches unity, leading to a significantly higher thermal conductivity than an actual porous structure.

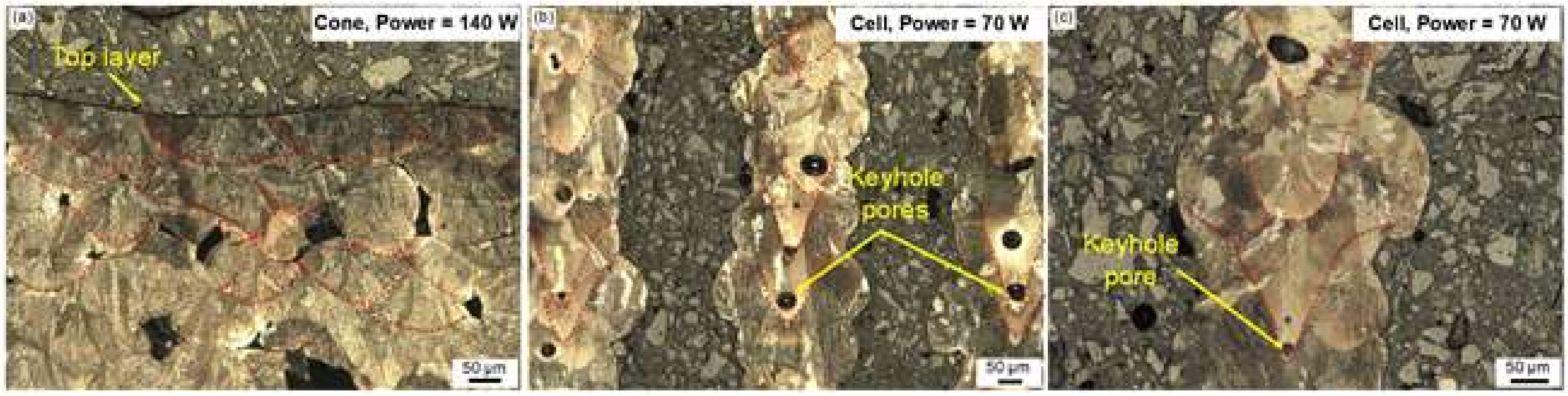

**Fig. 10. OM images of (a) a cone (bulk) fabricated at 140 W, and (b-c) S0.2-0.5-0.5 cellular structure produced at 70 W, showing melt pool boundaries (selectively delineated using dashed lines) and formation of keyhole pores in S0.2-0.5-0.5.**

*3.3.2. Pore size and porosity in S0.2-0.5-0.5*

SEM images from a large area of as-LPBF S0.2-0.5-0.5 specimens were used to roughly estimate pore sizes by averaging equivalent circular diameters. The data obtained are plotted against VED and $E_l$ in Fig. 11a, showing that pore equivalent diameters were ~ 100 µm.

The density of S0.2-0.5-0.5 and porosity for a range of VED and $E_l$ are shown in Figs. 11b and c, respectively, revealing porosity of > ~50-60% in all specimens. Pore sizes and porosity remained in the same ranges with increasing VED and $E_l$ (Figs. 11a and c), perhaps, owing to an insignificant increase in strut diameters (Fig. 8) and uncertainty in measurements.

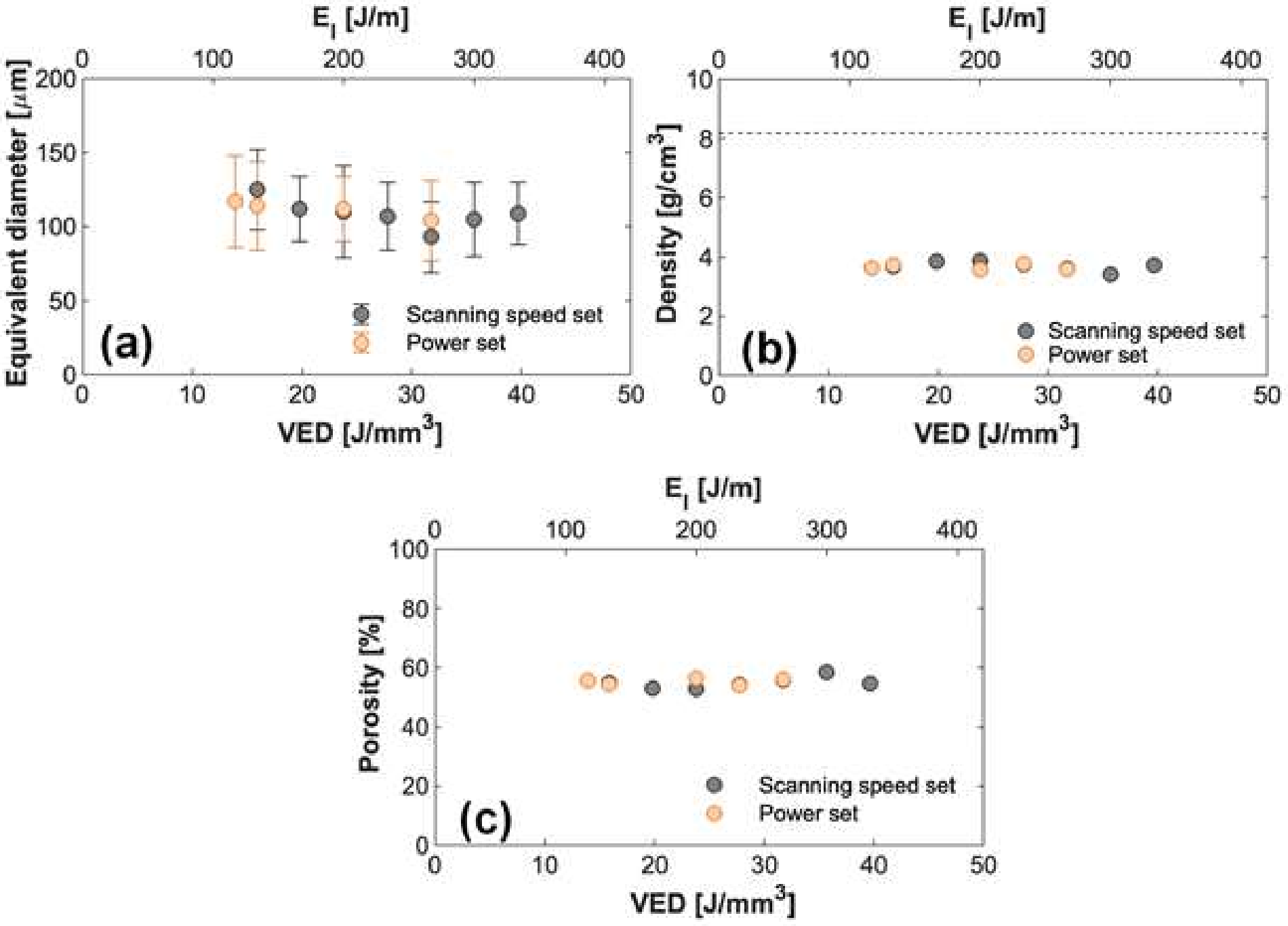

**Fig. 11 (a) Equivalent circular diameters of pores, (b) density of porous S0.2-0.5-0.5, and (d) porosity against VED and *El* varied by power and scanning speed, showing pore sizes of mostly ~100 µm, and porosity of > ~ 50 – ~ 60%. The dashed line in (b) marks the density of a bulk In718 (8.19 g/cm$^3$).**

*3.4. Electrochemical testing*

Nickel-based porous electrodes are often used in alkaline water electrolysers, a prominent electrochemical technology for the sustainable manufacturing of hydrogen [63]. To assess the potential of the LPBF-fabricated porous electrodes in the present work for this application, HER and OER were examined before and after nickel deposition. Only the sample S0.2-0.5-0.5 was tested as the optimum printable design.

Figs. 12a-d depict SEM micrographs of the nickel coatings on the In718 substrate at different magnifications. Fig. 12a shows a network of pores with an uneven size distribution (micrometer to submicron range), perhaps, owing to the macrostructural nonuniformity originated from partially melted particles attached to the struts in the additively manufactured In718 substrate. During the electrodeposition, gas ($H_2$) bubbles are produced as a result of the proton reduction. These bubbles

function as a dynamic template for the electrodeposition, aiding in the formation of finer pores of the deposited catalytic layer, as detailed elsewhere [54]. However, the partially melted particles with different sizes on the struts might have changed the bubble break-off diameters, resulting in a network of pores with nonuniform coverage, thickness, sizes and distribution. Furthermore, other factors such as nucleation, growth, and concentration of supporting electrolyte (and therefore rate of the hydrogen evolution used in templating) may influence the surface morphology of these coatings [54, 64]. Moreover, the side walls of the newly formed pores had a cauliflower-like structure (Fig. 12b-d). Point EDS analysis (Fig. S1) conducted on different locations of the coatings also revealed that the coatings contained > 99% nickel, indicating the high purity of the catalytic layer. XRD analysis confirms the presence of crystalline phases of nickel (Figs. S3a and b) with a cubic crystal system (COD card No. 96-151-2527).

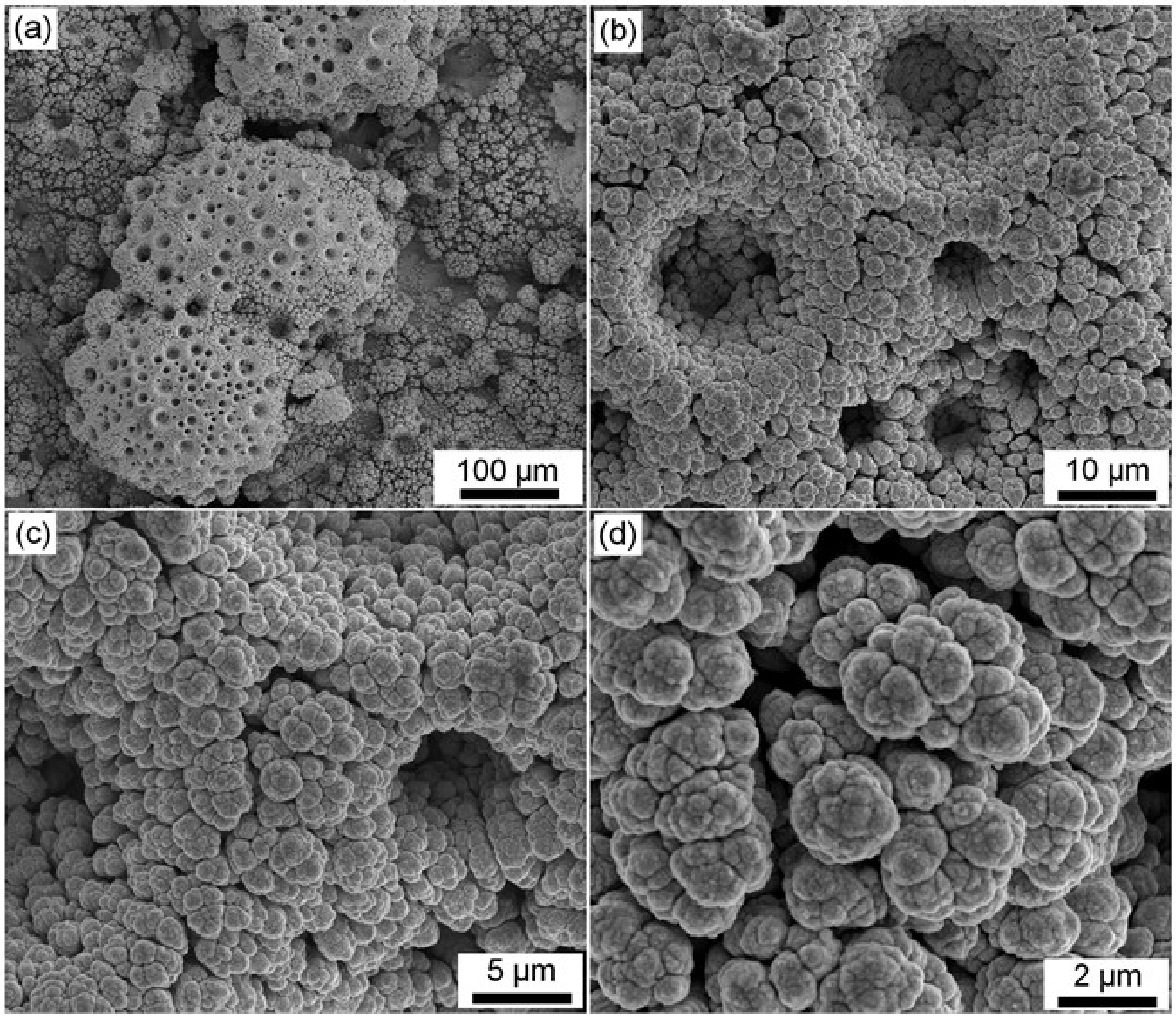

**Fig. 12. SEM micrographs of the surface of S0.2-0.5-0.5 after nickel electrodeposition at different magnifications, showing (a) a network of nonuniformly distributed pores, (b-d) closeups of fine pores (of the order of µm) at different locations.**

Prior to electrochemical performance for HER and OER, both bare and nickel-coated In718 were subjected to cyclic voltammetry at a scan rate of 50 mV/s in the extreme potential window of −1 to 2.3 V vs RHE (Fig. S4). A peak associated with hydrogen desorption is detected at 0.2 to 0.3 V vs RHE, while other peaks in the range of 1.0 V to 1.48 V vs RHE were attributed to hydroxide production ($M(OH)_x$) and subsequent oxidation to oxyhydroxides (α-/β-/γ- $M_{(x-1)}OOH$), favoring oxygen adsorption and reduction (Fig. S4) [65-67]. The maximum observed current for HER reached ~ −400 mA at −1 V vs RHE, whereas ~ 350 mA was obtained at −2.3 V vs RHE for OER. Generally, in alkaline conditions (at pH ≈ 14) HER starts around −0.826 V vs RHE and OER around 0.4 V vs RHE, while we observed a shift to less negative potentials on this theoretical onset value (~ −0.2 V vs RHE for HER and ~1.5 V vs RHE for OER) due to non-idealities introduced by the setup geometry and thin passivation layers generated by the background electrolyte, in accordance with Bui et al. [68]. An average value of open circuit voltage of ~ −800 mV was observed in our three electrode system. The corresponding applied potential was corrected with the iR compensation stemming from the ohmic resistance of the electrochemical cell, as explained above. Impedance spectra of both bare and nickel-coated In718 are presented as Nyquist plots (Figs. S5a and b) to obtain the ohmic resistance. A two-time constant parallel model has been employed to precisely simulate the impedance behaviour of electrodes (Fig. S6). Obtained values of current were normalised by the geometric surface area to measure current density ($j$, mA/cm$^2$). The geometric surface area was kept constant (1 cm$^2$) for all the electrodes. The corrected potential ($E_{iR\ corrected}$) was plotted against the corresponding current density (i.e., polarisation curve) for HER and OER, as shown in Figs. 13a and b, respectively. In the case of HER performance, the nickel-coated electrodes performed substantially better than the as-LPBF (bare) electrodes. More specifically, the nickel-coated electrode resulted in a larger maximum current density of ~ −372 mA/cm$^2$ at a less negative potential of ~ −0.4 V vs RHE compared to the bare electrode with a maximum current density of ~ −240 mA/cm$^2$ at ~ −0.6 V, indicating the superior performance of the former. This increase in the performance could be due to the higher surface area of Ni-coated electrode. An overpotential (η) of 150 mV (Table S2 and S4) was required to derive ~−10 mA in the nickel-coated In718, whereas in bare In718, an overpotential of ~400 mV was required to derive the same current (Fig S7a). In order to access the activity of the sample, Tafel plots were derived from the polarisation curve in 1.0 M KOH. The Tafel plots showed an overpotential dependent behaviour (Figs. S8a and b). In the low overpotential regime of HER, a Tafel slope of 122 mV/dec was obtained, which is closer to the values in the literature, whereas in the mid-higher overpotential regime, the value of Tafel slope was 243 mV/dec (Fig. S8a). This anomaly from the simple linearity could be due to several factors, such as back reaction at low overpotentials, mass transport together with the blocking effect of produced $H_2$ bubbles at high overpotentials, formation of a large number of N–H moieties, and the dependence of adsorbed hydrogen intermediate on overpotential [69-72]. Similar trend in Tafel plot was observed for OER with Tafel slopes of 73.74 and 291 mV/dec in low and high overpotential regimes, respectively (Fig S8 b).

In OER, bare and nickel-coated electrodes exhibited similar performance, both showing a maximum current density of ~350 mA/cm$^2$ at ~ 1.8 V vs RHE (Fig. 13b) with an overpotential of 332 mV at 10 mA (Table S3, S5 and Fig. S 7b). These results could be explained by two factors: first, complete coverage of the surface by nickel coatings, and, second, the presence of iron in In718 as a major alloying element. The higher activity and surface area of nickel coatings could lead to the better performance of the nickel-coated electrodes for HER. However, for OER, iron (Fe) in as-LPBF samples could contribute to obtaining a current density as high as that obtained from the nickel-coated samples, despite a much lower surface area in the former. It has been shown that iron facilitates oxygen production by reducing the oxygen overpotential, thus improving OER performance [73-76]. Xiao et al. investigated the function of iron and nickel in OER. They explained the mechanism of OER using quantum mechanics (density functional theory, DFT). They demonstrated the iron-nickel synergy, in which $d^4$ Fe(IV) led to the creation of an active O radical intermediate, followed by O-O coupling by Ni(IV), resulting in improved performance [77]. Further, a review by Anantharaj et al. [76] summarises the observations from recent works explaining the mechanistic role of iron, either in nickel or cobalt matrices for OER. The better conductivity of catalyst is assumably connected to the electrocatalytic activity and assumed to play a role in overall performance [76]. Boettcher and co-workers [74] found that the effective conductivity of Ni(OH)2/NiOOH thin film electrode increased by more than 30 fold with iron incorporation. However, these findings cannot explain the rationale for increased activity on their own. Rong's latest research [78] revealed the reactive location of iron in nickel hydroxide matrices. According to them, iron species found on the outer surface region had an important part in acting as OER active sites and increasing overall performance. Herein, iron in the nickel-coated samples was not easily accessible, and as a consequence, would not play a significant role in the catalytic activity towards OER, although the high surface area of the coatings compensated for that. On the other hand, in the as-LPBF samples, iron contribution would be much more significant, but the current density was not higher than the coated ones, probably, owing to their lower surface area. To access the stability and durability of the nickel-coated In718, a chronopotentiometeric test was performed. A constant current of 400 mA was applied for 24 h and corresponding potential response was recorded (Fig. S9). Despite a few spikes in the potential transient which could be due to high bubble formation, the electrode exhibited a good stability.

Table 3 compares the results above with those reported by others under similar testing conditions. As seen, they are comparable to the state-of-the-art nickel-based materials for HER and OER, suggesting promising performance of the electrodes in the present work. Future work will focus on the stability and durability check of the nickel-coated electrodes by means of accelerated stress testing for longer durations. This can motivate future works, including electrochemical testing in flow cells, further modification of the designs, and optimisation of chemical compositions of substrates and coatings, to find the best combination of unit cell geometries, coatings and substrate materials with the intent to maximise the obtained overall performance towards HER and OER.

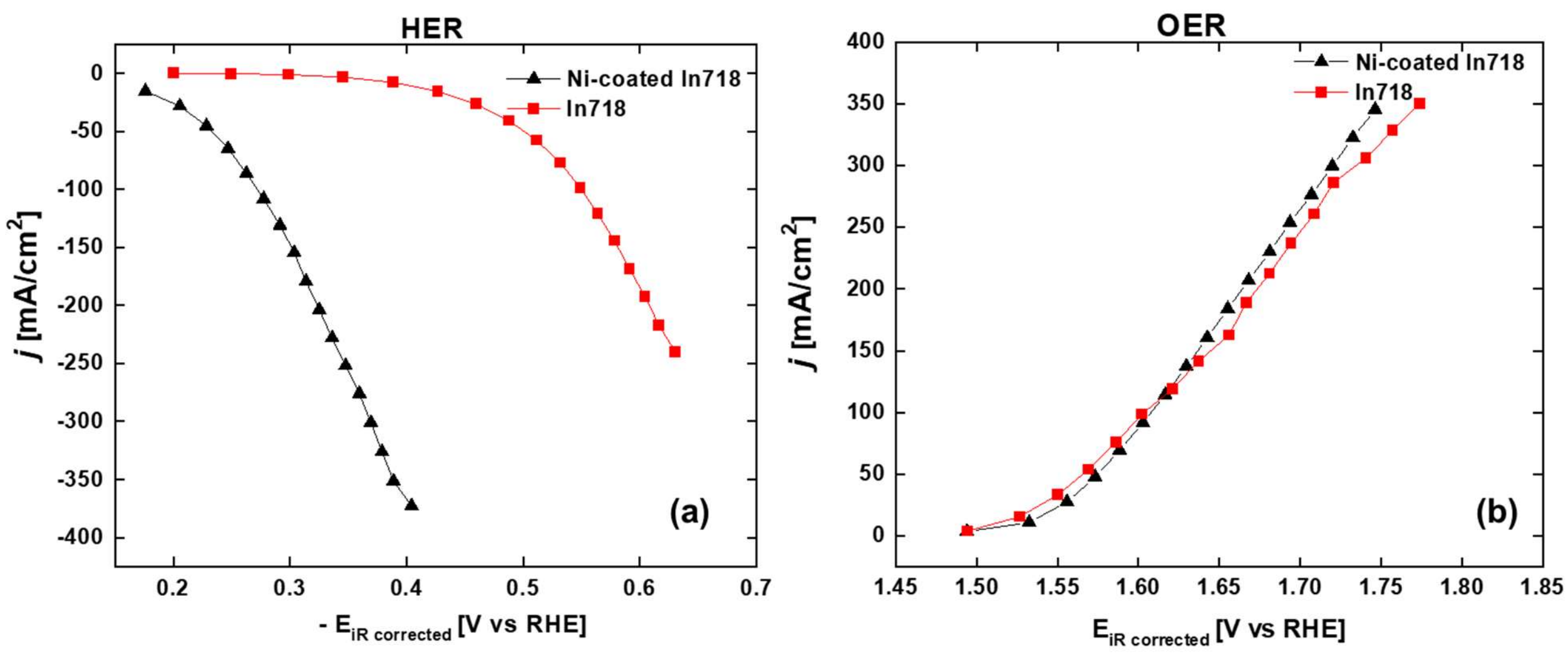

**Fig. 13. Polarisation curves for (a) HER and (b) OER obtained from nickel-coated In718 (black) and as-LPBF In718 (red) electrodes measured in 1 M KOH.**

**Table 3. Comparative performance of various nickel-based materials for HER and OER in alkaline environment (1 M KOH) in 3 electrode cell configuration. Potentials are represented vs RHE, unless otherwise specificed.**

| Material | Electrolyte | Reactions | $j$ (mA/ $cm^2$ ) | Potential (V vs RHE) | $\eta_{10}$ mV | $\eta_{100}$ mV | Tafel slope mV/dec | Reference |
|---|---|---|---|---|---|---|---|---|
| Ni-coated In718 | 1 M KOH | HER | −360 | −0.4 | 150 | 257 | 122, 243 | This work |
| Ni-coated In718 | 1 M KOH | OER | 350 | 1.8 | 332 | 403 | 73.74, 291 | This work |
| NixCux on 3D printed PLA | KOH (concentration not reported) | HER | −15 | −1.4 vs Ag/AgCl 3M KCl | -- | -- | -- | [79] |
| NixCu2x on 3D printed PLA | KOH (concentration not reported) | HER | −27 | −1.4 vs Ag/AgCl 3M KCl | -- | -- | -- | [79] |
| NixCu3x on 3D printed PLA | KOH (concentration not reported) | HER | −40 | −1.4 vs Ag/AgCl 3M KCl | -- | -- | -- | [79] |
| Ni-CoO | 1 M KOH | HER | −300 | −0.31 | -- | -- | 59 | [80] |
| Ni-Zn-CoO | 1 M KOH | HER | −300 | −0.24 | 53 | -- | 47 | [80] |
| $Ni_3N$/Ni/NF | 1 M KOH | HER | −100 | −0.126 | 12 | 64 | 29.3, 80.1 | [69] |
| Ni-ABS (acrylonitrile butadiene styrene) 3D | 1 M KOH | HER | −200 | −0.26 | 170 | -- | 94.3 | [81] |
| NiW-ABS (acrylonitrile butadiene styrene) 3D | 1 M KOH | HER | −200 | −0.16 | 198 | -- | 105.2 | [81] |
| NiFe-ABS (acrylonitrile butadiene styrene) 3D | 1 M KOH | OER | 100 | 1.45 | 370 | -- | 57 | [81] |
| 3D-printed nickel-PLA | 1 M NaOH | HER | −100 | −0.9 | -600[a] | -- | 105 | [8] |
| NFNS@NiP@Truss | 1 M KOH | HER | −50 | −0.35 | 243 | -- | 141 | [82] |
| NFNS@NiP@Truss | 1 M KOH | OER | 100 | 1.48 | 197 | -- | 51 | [82] |
| 3D-printed nickel-PLA | 1 M NaOH | OER | 150 | 2.5 | 800[a] | -- | 157 | [8] |
| $Ni_xFe_yOH_z$/Ni-coated on 3D printed ABS mesh | 1 M KOH | OER | 100 | 1.55 | 250 | 300 | 53 | [83] |
| Graphene-Ni-Fe modified 3D printed fused filament sheets | 0.1 M KOH | OER | 10 | 1.6 | 519 | -- | 46 | [84] |

[a] These values for HER and OER overpotentials are given for a current density of 50 mA/cm$^2$.

## 4. Conclusions

Thin porous electrodes were produced using LPBF to construct strut-based lattice structures with a design thickness ranging from 400 µm to 1 mm. These electrodes consisted of BCC unit cells with dimensions of 200–500 µm and strut diameters of 100 and 200 µm. Subsequent to the LPBF process, a porous nickel catalyst layer was deposited on the electrodes to enhance electrochemical performance. Several key findings emerged from this study, providing a foundation for future research and advancements in LPBF technology for energy applications. Notably, the investigation revealed challenges, such as excessive melting and pore closure, when utilising unit cells of 200 µm. The optimum design identified a thickness of 500 µm, unit cells of 500 µm, and strut diameters of 200 µm, with variations in laser energy density having minimal impact on porosity and pore sizes. Deeper melt pools in lattice structures attributed to lower thermal conduction rates, resulted in the absence of large unmelted/sintered areas observed in bulk structures. Electrodeposition of nickel created a pore network with fine pores in the microns and submicron range, complementing the ~100 µm pores achieved through LPBF. Electrochemical tests demonstrated enhanced performance for the nickel-coated electrode in HER compared to the bare specimen. Both electrode types, however, exhibited similar performance in OER. The combination of LPBF of In718 and nickel electrodeposition proved effective for producing porous electrodes, albeit as a proof-of-concept, emphasising the need for further in-depth studies and technological modifications before commercialisation.

**Author contributions**

**A. Zafari:** Methodology, Formal analysis, investigation, Writing-Original draft, Project administration; **K. Kiran:** Conceptualisation; Methodology, Formal analysis, investigation, Writing-Original draf; **I. Gimenez-Garcia**: Conceptualisation, investigation, Writing- Review & Editing; **K. Xia:** Writing- Review & Editing, Resources; **I. Gibson:** Supervision, Writing-Review & Editing; **D.**

**Jafari:** Conseptulisation, Supervision, Funding acquisition, Project administration, Writing-Review & Editing.

## Acknowledgements

This work was supported by the European Space Agency (ESA) under the scheme Discovery Program – Early Technology Development (Project No. 4000135470/21/NL/GLC/ov). The views expressed herein can in no way be taken to reflect the official opinion of ESA and are not intended to endorse particular technologies, companies, or products. The authors acknowledge the contribution of Dr. A. Forner-Cuenca to conceptualisation, funding acquisition, and editing the manuscript. I.G.G gratefully acknowledges funding through the Postgraduate Fellowships program from "La Caixa" foundation (ID 100010434, 808 fellowship code LCF/BQ/EU20/11810076). We appreciate the facilities and technical assistance provided by the Bio21 Ian Holmes Imaging Centre at the University of Melbourne, Australia, and the $MS^3$ Microscopy/Preparation Laboratory at the University of Twente, the Netherlands.

## Data availability statement

The raw/processed data required to reproduce these findings cannot be shared at this time due to technical or time limitations.

# Framework for additive manufacturing of porous Inconel 718 for electrochemical applications

## Supplementary Information

Ahmad Zafari[a,1*], Kiran Kiran[b,1], Inmaculada Gimenez-Garcia[b], Kenong Xia[c], Ian Gibson[a], Davoud Jafari[a,*]

[a]*Department of Design, Production and Management, Faculty of Engineering Technology, University of Twente, 7500AE, The Netherlands*

[b]*Department of Chemical Engineering and Chemistry, Eindhoven University of Technology, 5600MB, Eindhoven, The Netherlands*

[c]*Department of Mechanical Engineering, The University of Melbourne, VIC 3010, Australia*

[1]These authors contributed equally to the work.

*** Corresponding authors:** ahmad.zafari@utwente.nl; davoud.jafari@utwente.nl

# Table of Contents

## 1. Working parameters for deposition

**Table S1. Working conditions for electrodeposition represented in tabular format.**

| Working conditions | |
|---|---|
| Metal Precursor | Nickel (II) chloride hexahydrate (0.1 M) |
| Supporting electrolyte | Ammonium chloride, $NH_4Cl$ (1.5 M) |
| Current density ($I_{dep}$) | -1 A / $cm^2$ |
| Deposition time ($T_{dep}$) | 100 s |
| Working & counter electrode | In718 and Pt mesh |

## 2. EDS and XRD analysis

### 2.1 Point EDS

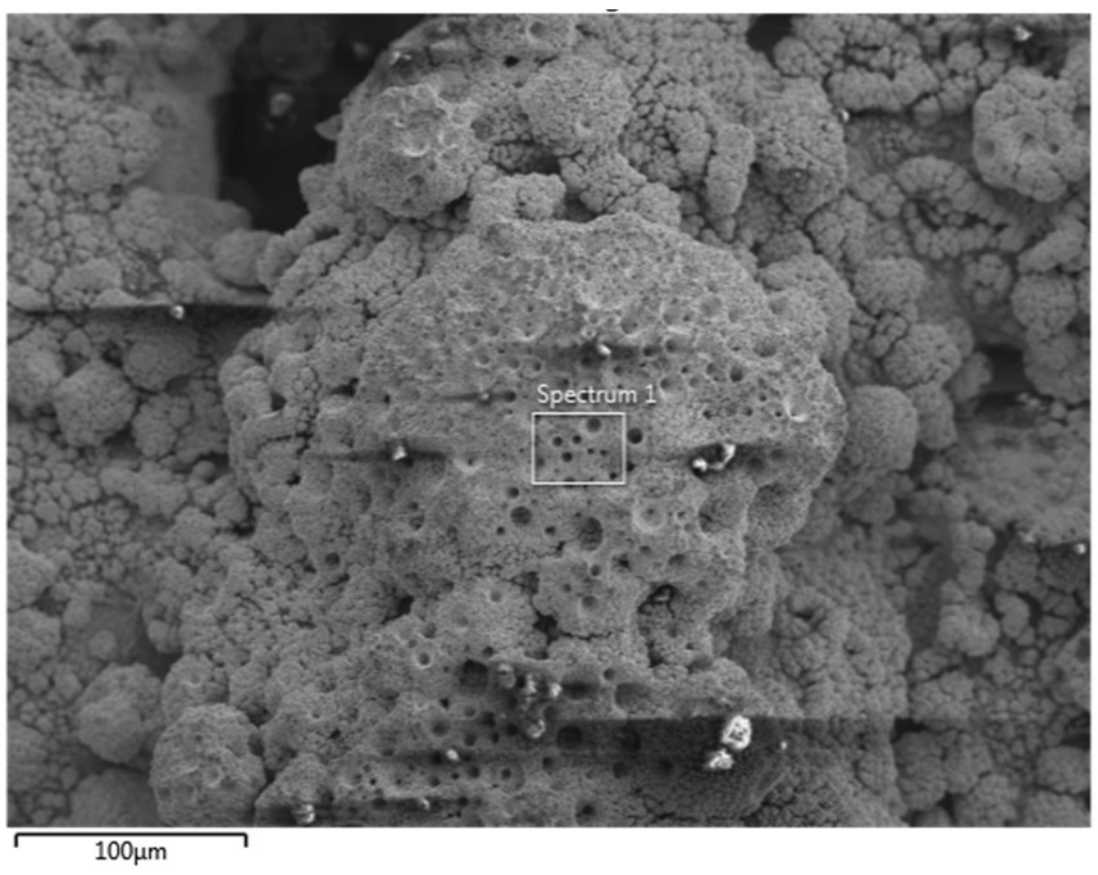

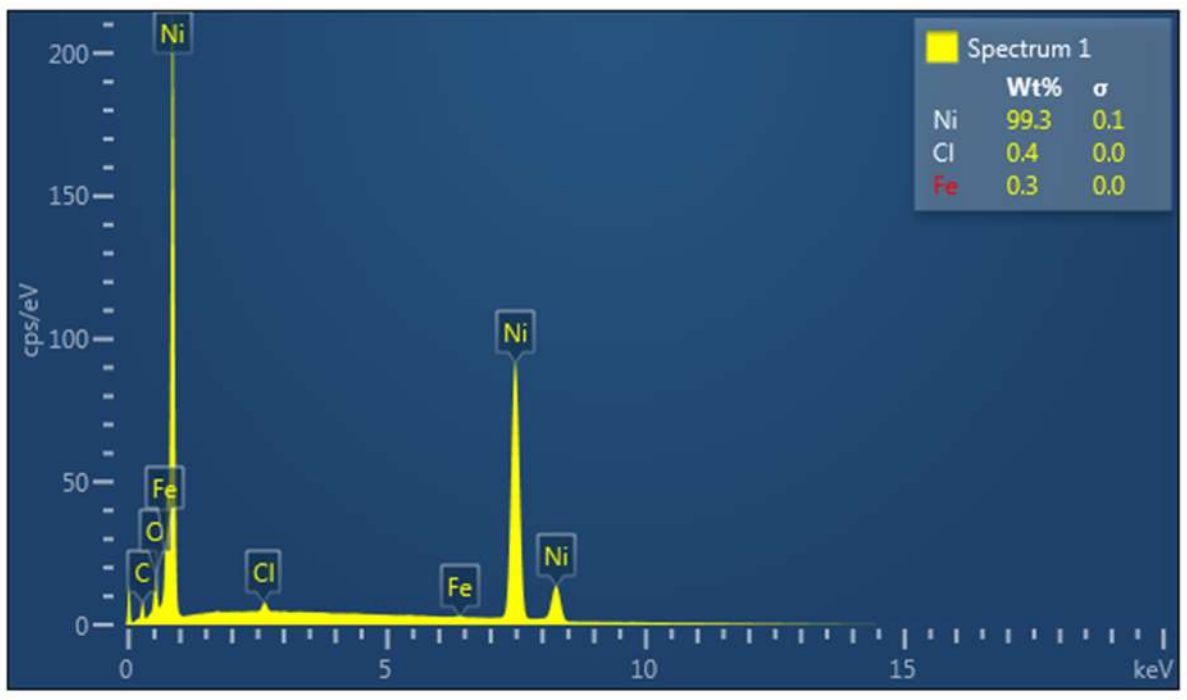

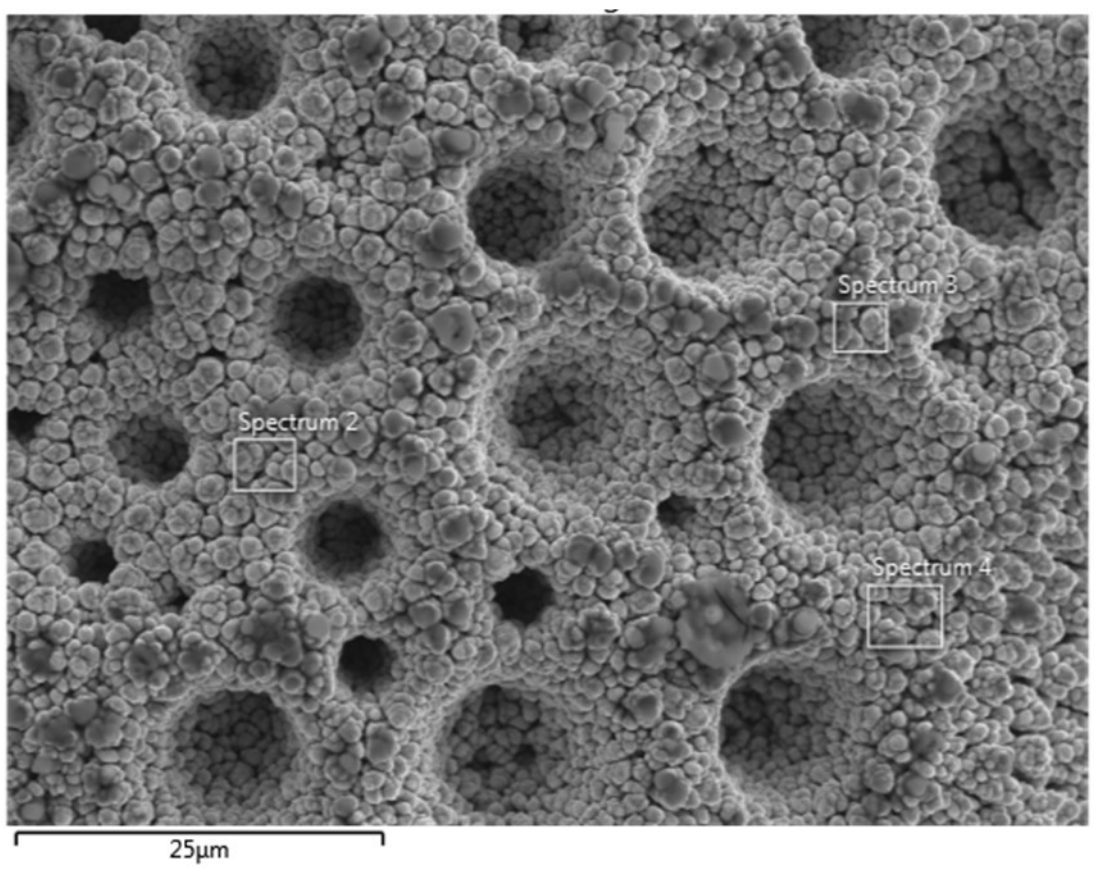

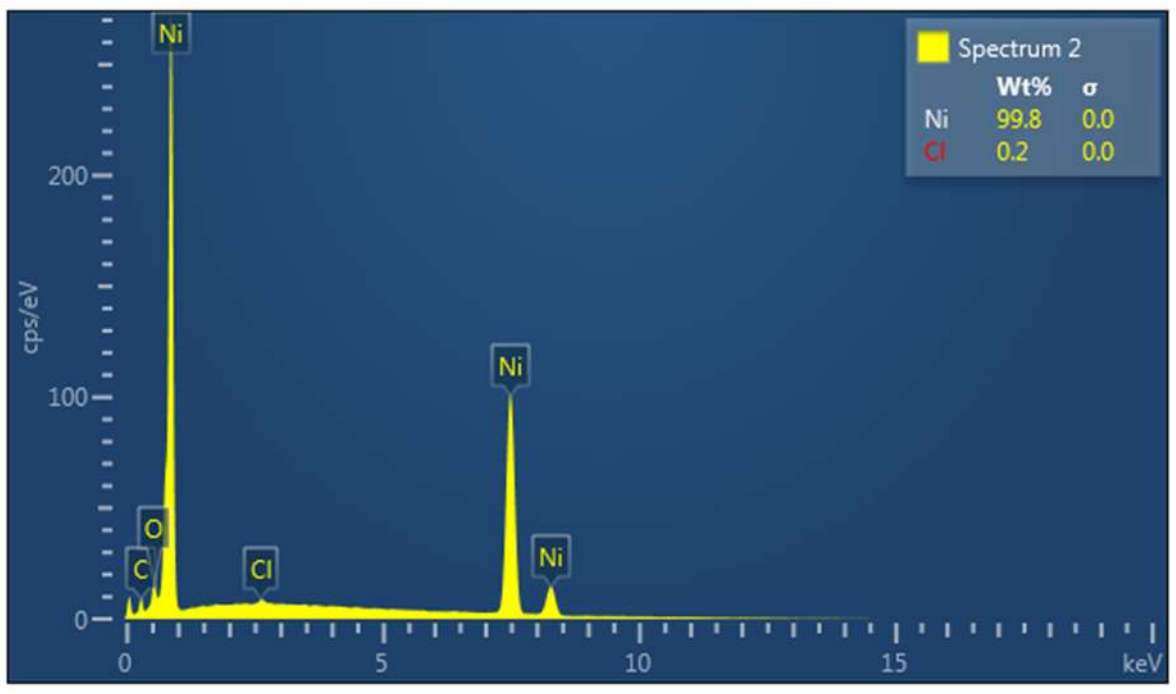

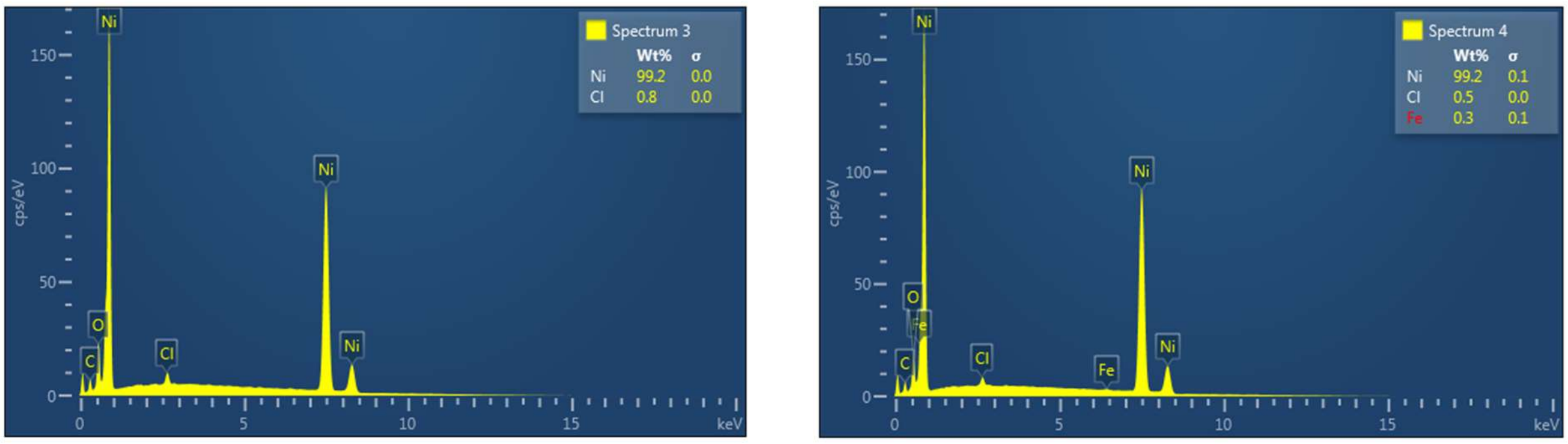

**Fig S1. Point EDS analysis conducted on different locations of nickel coatings, showing >99% nickel in the coatings.**

## 2.2 EDS map

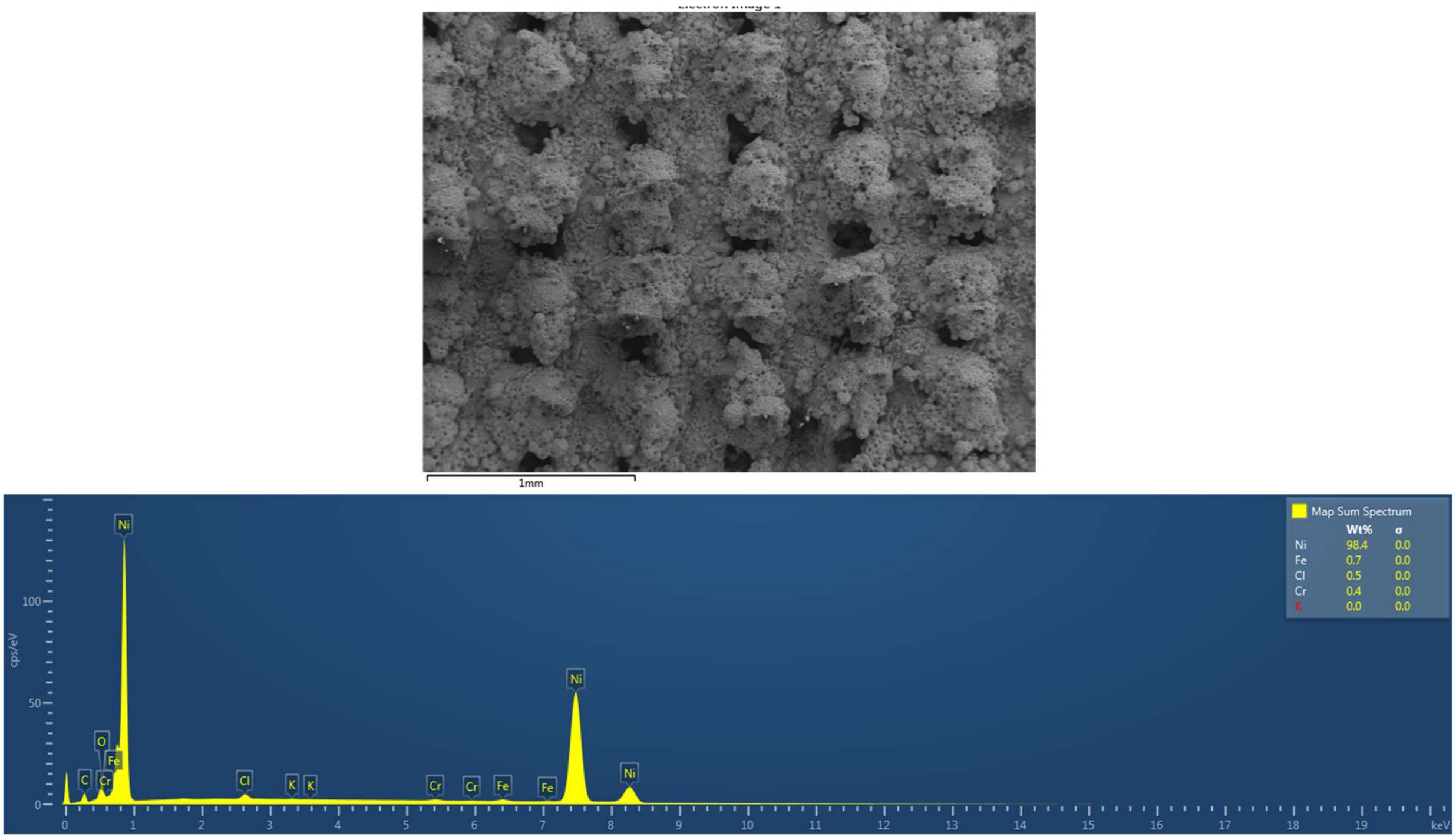

**Fig S2. EDS map form a 3mm × 3 mm area (top) and the map sum spectrum achieved from this area (bottom), revealing > 98% nickel from the coatings. The remaining elements detected are from the In718 substrate.**

## 2.3 XRD

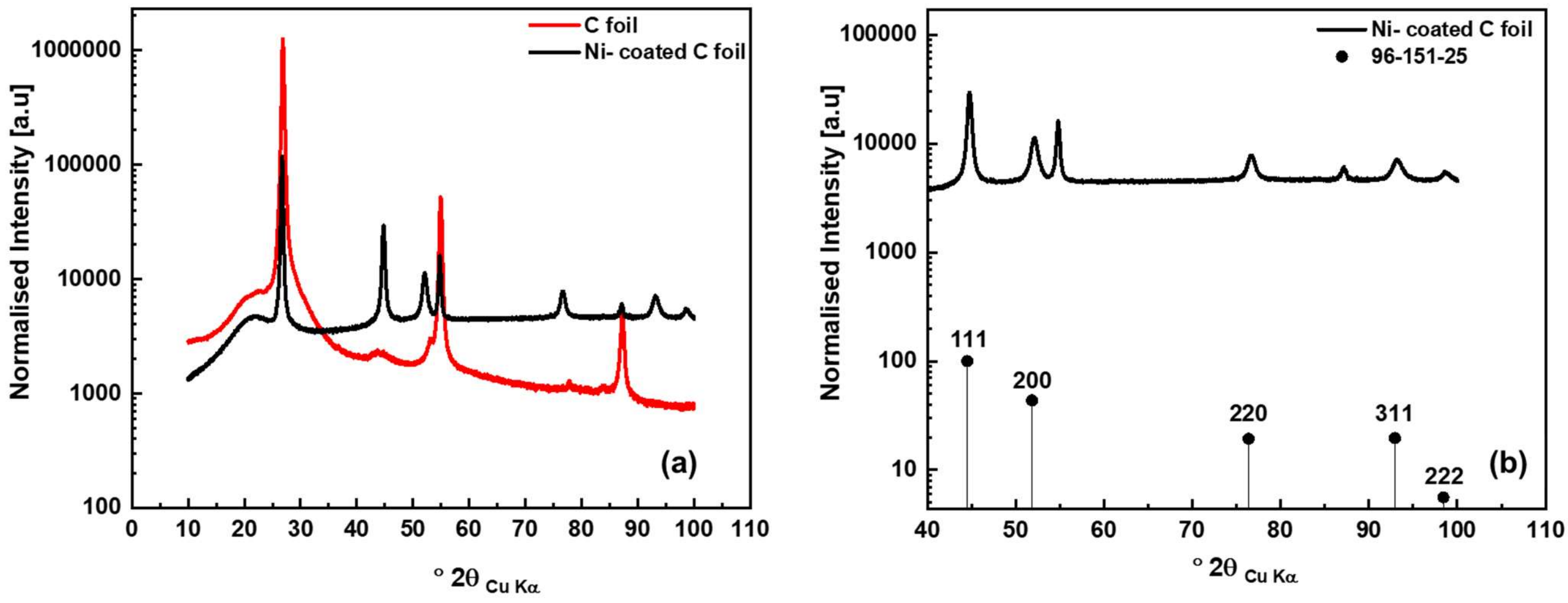

**Fig S3. XRD analysis of (a) graphite (C) foil and Ni-coated graphite (C) foil; (b) zoomed in version of Ni-coated graphite (C) foil. Intensities are presented as logarithmic scale.**

## 3. Cyclic voltammetry

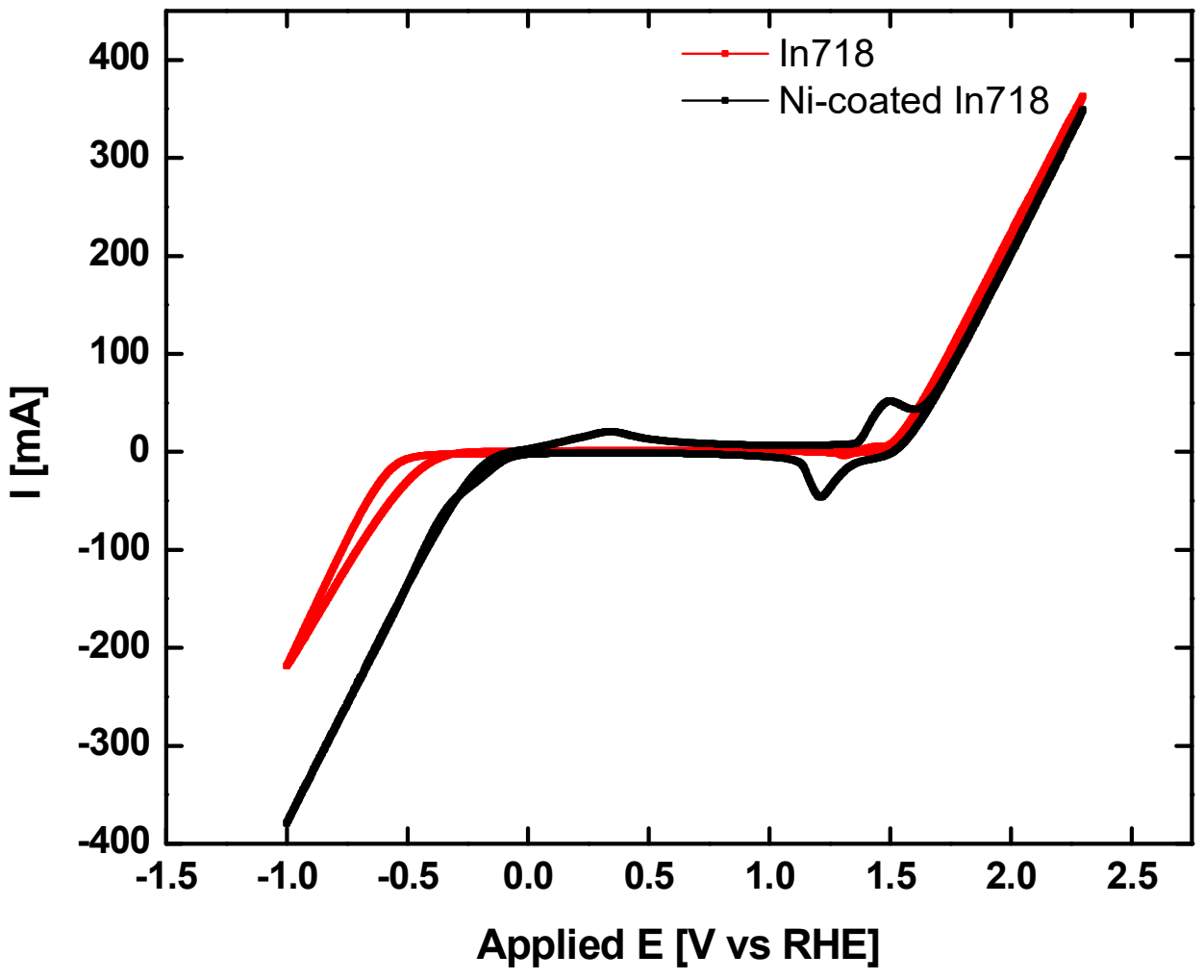

**Fig S4. Cyclic voltammetry of bare and nickel-coated In718 at 50 mV/s in 1 M KOH.**

## 4. Impedance measurement by EIS and equivalent circuit

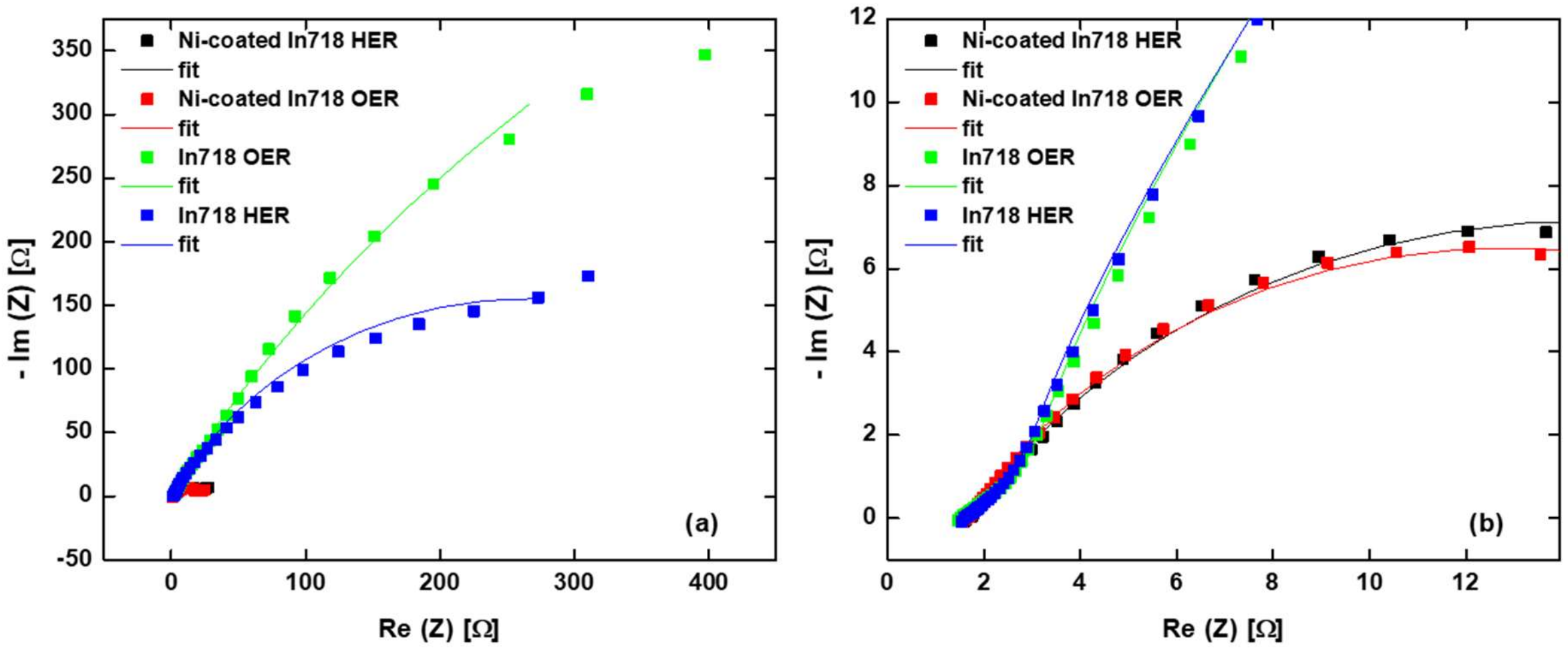

**Fig S5. (a) Impedance spectra of bare and nickel-coated In718 in the frequency range of 300 KHz to 100 mHz, (b) zoomed in scale is shown showing ohmic resistance. Markers are representing the experimental value and lines are representing the fit.**

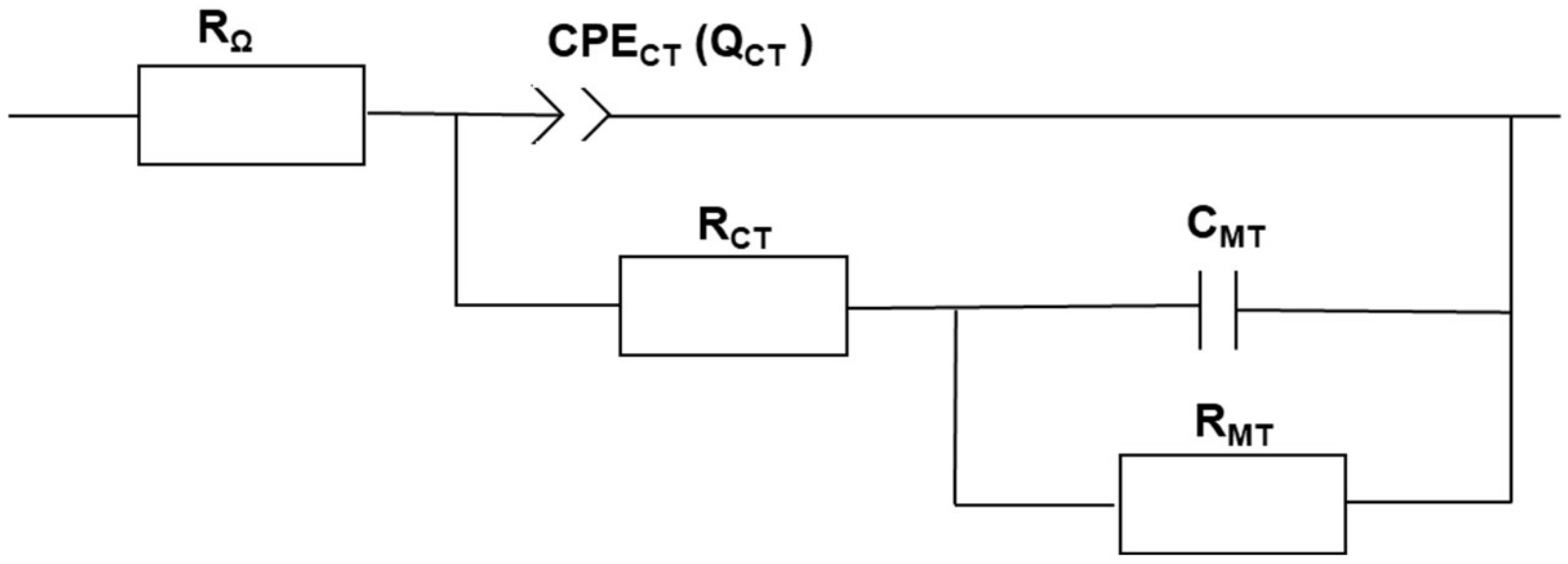

**Fig S6. Equivalent circuit used for fitting the EIS impedance spectra.**

To accurately define and simulate the impedance behaviour of various types of electrodes [65, 85-87], a two-time constant parallel model has been employed. It consists of $R_{\Omega}$, which quantifies the system's ohmic response, $R_{CT}$, and constant phase element ($CPE_{CT}$), which construct a first time constant connected to charge-transfer kinetics. The CPE accounts for capacitance dispersion, which can cause distortion. Adsorption of molecules on the surface of the electrodes is hindered, resulting in a second time constant specified by $C_{MT}$ and $R_{MT}$, particularly for the system's mass transport-related impedances.

## 5. Calculation of reaction overpotentials

For all overpotential calculations, an average open circuit voltage (OCV) of 0.8 V was used. Electrochemical impedance spectroscopy (EIS) was used to determine the ohmic losses of each test. A resistance of 1.6 ohm [Ω] was measured for HER for the Ni-coated In718 sample. To calculate this iR compensation, the current density [A $cm^{-2}$] was multiplied by value of the resistance (1.6 Ω). The obtained iR correction was then subtracted from the applied potential, E[V], to derive the iR-corrected potential value, $E_{iR\ corrected}$ (Fig 13). We added the recorded OCV value to the iR-corrected potential to obtain $E^*$ (after OCV correction). The thermodynamic potential of HER and OER at the operating pH level of 14 (−0.826 V and 0.4 V respectively) was subtracted from $E^*$ to find the reaction overpotential, η[V]. For simplicity, the data in the following tables has been rounded upto three significant figures.

### 5.1 Hydrogen evolution reaction overpotentials for the as-printed In718 porous structure

**Table S2. Experimental data and calculation of overpotential of the as-printed In718 substrate for HER.**

| E [V] | −*j* [mA/ $cm^2$ ] | iR calculation with 1.54 Ω | $E_{ir\text{-}corrected}$ [$V_{RHE}$] =E-iR | E* after adding OCV [$V_{RHE}$] = $E_{iR}$+OCV | Overpotential [V] = $E^*$-$E_{thermodynamic}$ |
|---|---|---|---|---|---|
| −0.201 | 0.115 | −0.000178 | −0.200 | −1.00 | −0.174 |
| −0.250 | 0.389 | −0.000599 | −0.250 | −1.05 | −0.224 |
| −0.301 | 1.22 | −0.00187 | −0.299 | −1.10 | −0.273 |
| −0.350 | 3.41 | −0.00525 | −0.345 | −1.15 | −0.319 |
| −0.401 | 7.93 | −0.0122 | −0.388 | −1.19 | −0.362 |
| −0.451 | 15.5 | −0.0239 | −0.427 | −1.23 | −0.401 |
| −0.501 | 26.6 | −0.0409 | −0.460 | −1.26 | −0.434 |
| −0.551 | 40.8 | −0.0629 | −0.488 | −1.29 | −0.462 |
| −0.600 | 57.9 | −0.0891 | −0.511 | −1.31 | −0.485 |
| −0.650 | 77.3 | −0.119 | −0.532 | −1.33 | −0.506 |
| −0.701 | 98.7 | −0.152 | −0.549 | −1.35 | −0.523 |
| −0.751 | 121 | −0.187 | −0.564 | −1.36 | −0.538 |
| −0.800 | 144 | −0.222 | −0.578 | −1.38 | −0.552 |
| −0.851 | 168 | −0.260 | −0.591 | −1.39 | −0.565 |
| −0.900 | 192 | −0.296 | −0.604 | −1.40 | −0.578 |
| −0.951 | 217 | −0.334 | −0.616 | −1.42 | −0.590 |
| −1.00 | 240 | −0.370 | −0.630 | −1.43 | −0.604 |

### 5.2 Oxygen evolution reaction overpotentials for the as−printed In718 porous structure

**Table S3. Experimental data and calculation of overpotential of the as-printed In718 substrate for OER.**

| E [V] | $-j$ [mA/ cm$^2$ ] | iR calculation with 1.5 Ω | $E_{ir\text{-}corrected}$ [$V_{RHE}$] =E-iR | E* after adding OCV [$V_{RHE}$] = $E_{iR}$+OCV | Overpotential [V] = $E^*$- $E_{thermodynamic}$ |
|---|---|---|---|---|---|
| 1.50 | 3.86 | 0.00578 | 1.49 | 0.694 | 0.294 |
| 1.55 | 15.6 | 0.02343 | 1.53 | 0.727 | 0.327 |
| 1.60 | 33.3 | 0.0500 | 1.55 | 0.750 | 0.350 |
| 1.65 | 53.9 | 0.0809 | 1.57 | 0.769 | 0.369 |
| 1.70 | 75.8 | 0.114 | 1.59 | 0.786 | 0.386 |
| 1.75 | 98.7 | 0.148 | 1.60 | 0.802 | 0.402 |
| 1.80 | 119 | 0.179 | 1.62 | 0.821 | 0.421 |
| 1.85 | 142 | 0.213 | 1.64 | 0.837 | 0.437 |
| 1.90 | 163 | 0.244 | 1.66 | 0.856 | 0.456 |
| 1.95 | 189 | 0.283 | 1.67 | 0.867 | 0.467 |
| 2.00 | 213 | 0.319 | 1.68 | 0.881 | 0.481 |
| 2.05 | 237 | 0.356 | 1.69 | 0.894 | 0.494 |
| 2.10 | 261 | 0.392 | 1.71 | 0.909 | 0.509 |
| 2.15 | 286 | 0.429 | 1.72 | 0.921 | 0.521 |
| 2.20 | 306 | 0.459 | 1.74 | 0.941 | 0.541 |
| 2.25 | 329 | 0.493 | 1.76 | 0.957 | 0.557 |
| 2.30 | 351 | 0.526 | 1.77 | 0.974 | 0.574 |

## 5.3 Hydrogen evolution reaction overpotentials for the Ni DHBT-coated In718 porous structure

**Table S4. Experimental data and calculation of overpotential of the Ni-DHBT coated In718 substrate for HER.**

| E [V] | $-j$ [mA/ cm$^2$ ] | iR calculation with 1.6 Ω | $E_{ir\text{-}corrected}$ [$V_{RHE}$] =E-iR | E* after adding OCV [$V_{RHE}$] = $E_{iR}$+OCV | Overpotential [V] = $E^*$-$E_{thermodynamic}$ |
|---|---|---|---|---|---|
| −0.201 | 15.4 | −0.0247 | −0.176 | −0.976 | −0.150 |
| −0.251 | 28.3 | −0.0453 | −0.206 | −1.01 | −0.180 |
| −0.301 | 45.2 | −0.0724 | −0.228 | −1.03 | −0.202 |
| −0.351 | 64.8 | −0.104 | −0.247 | −1.05 | −0.221 |
| −0.401 | 86.3 | −0.138 | −0.263 | −1.06 | −0.237 |
| −0.451 | 108 | −0.173 | −0.277 | −1.08 | −0.251 |
| −0.501 | 130 | −0.209 | −0.291 | −1.09 | −0.265 |
| −0.551 | 154 | −0.247 | −0.304 | −1.10 | −0.278 |
| −0.600 | 179 | −0.287 | −0.314 | −1.11 | −0.288 |
| −0.651 | 204 | −0.326 | −0.325 | −1.13 | −0.299 |
| −0.701 | 228 | −0.365 | −0.336 | −1.14 | −0.310 |
| −0.751 | 252 | −0.403 | −0.348 | −1.15 | −0.322 |
| −0.801 | 276 | −0.441 | −0.359 | −1.16 | −0.333 |
| −0.851 | 301 | −0.481 | −0.369 | −1.17 | −0.343 |
| −0.901 | 326 | −0.522 | −0.379 | −1.18 | −0.353 |
| −0.951 | 351 | −0.562 | −0.389 | −1.19 | −0.363 |
| −1.00 | 373 | −0.597 | −0.404 | −1.20 | −0.378 |

## 5.4 Hydrogen evolution reaction overpotentials for the Ni DHBT-coated In718 porous structure

**Table S5. Experimental data and calculation of overpotential of the Ni-DHBT coated In718 substrate for OER.**

| E [V] | $j$ [mA/ cm$^2$ ] | iR calculation with 1.6 Ω | $E_{ir\text{-}corrected}$ [$V_{RHE}$] =E-iR | E* after adding OCV [$V_{RHE}$] = $E_{iR}$+OCV | Overpotential [V] = $E^*$- $E_{thermodynamic}$ |
|---|---|---|---|---|---|
| 1.50 | 3.53 | 0.00565 | 1.49 | 0.694 | 0.294 |
| 1.55 | 10.9 | 0.0175 | 1.53 | 0.732 | 0.332 |
| 1.60 | 27.4 | 0.0439 | 1.56 | 0.756 | 0.356 |
| 1.65 | 47.7 | 0.0764 | 1.57 | 0.773 | 0.373 |
| 1.70 | 69.6 | 0.111 | 1.59 | 0.788 | 0.388 |
| 1.75 | 91.8 | 0.147 | 1.60 | 0.803 | 0.403 |
| 1.80 | 114 | 0.183 | 1.62 | 0.816 | 0.416 |
| 1.85 | 138 | 0.220 | 1.63 | 0.830 | 0.430 |
| 1.90 | 161 | 0.257 | 1.64 | 0.843 | 0.443 |
| 1.95 | 184 | 0.294 | 1.66 | 0.855 | 0.455 |
| 2.00 | 207 | 0.332 | 1.67 | 0.868 | 0.468 |
| 2.05 | 230 | 0.369 | 1.68 | 0.881 | 0.481 |
| 2.10 | 254 | 0.406 | 1.69 | 0.894 | 0.494 |
| 2.15 | 277 | 0.443 | 1.71 | 0.907 | 0.507 |
| 2.20 | 300 | 0.480 | 1.72 | 0.920 | 0.520 |
| 2.25 | 323 | 0.517 | 1.73 | 0.933 | 0.533 |
| 2.30 | 346 | 0.553 | 1.75 | 0.946 | 0.546 |

## 6. Comparison of performance for hydrogen and oxygen evolution reaction as a function of reaction overpotential and current density

After correction for the intrinsic ohmic drop of the 3-electrode cell and the measured open circuit voltage, the performance for HER and OER for the studied pristine and coated samples can be seen in Fig S1 in terms of reaction overpotential, to allow easier comparison with other reaction set-ups or reference electrodes.

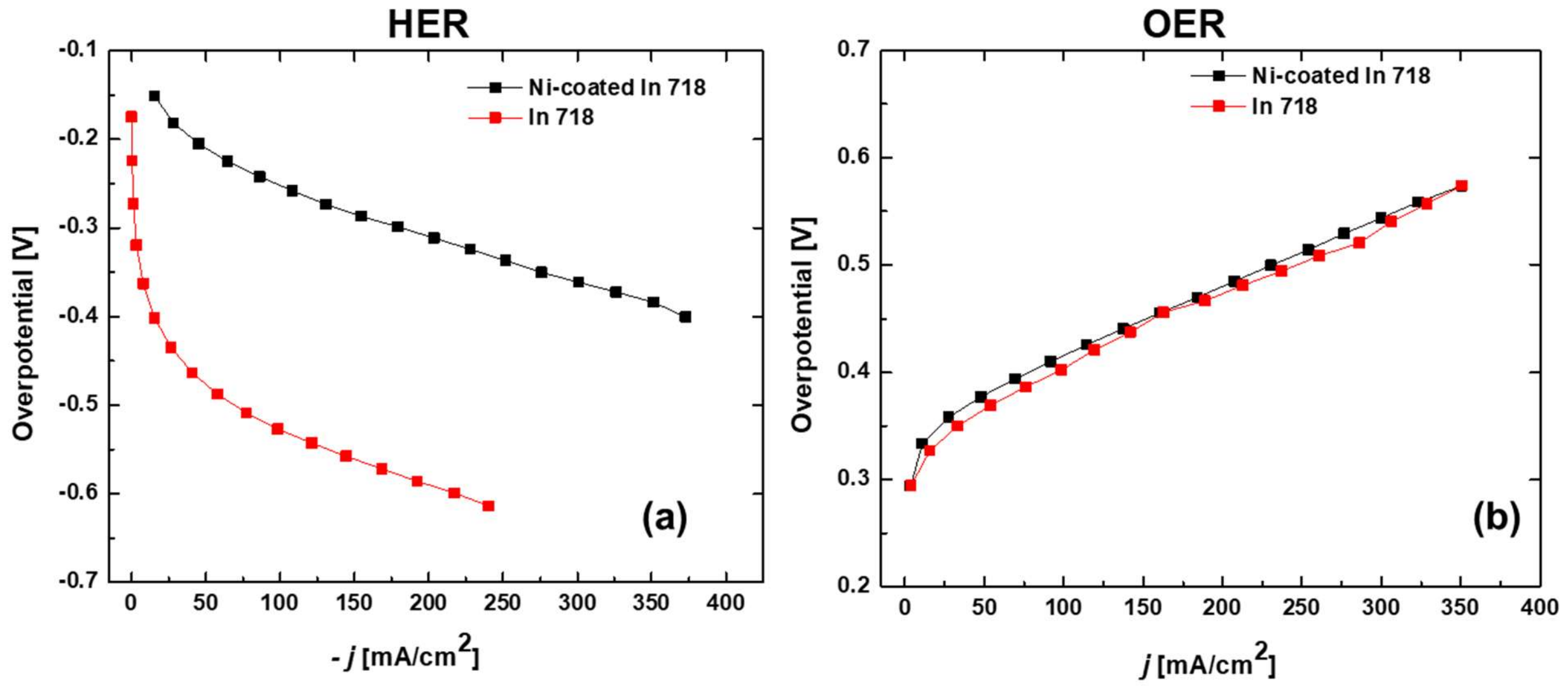

**Fig S7. Polarisation curves depicting overpotential as a function of current density for (a) HER and (b) OER obtained from Ni-coated In718 (black) and as-LPBF In718 (red) electrodes.**

## 7. Tafel slope for nickel-coated In718 for HER and OER respectively

The intrinsic activity of coated electrodes was evaluated on the basis of Tafel slope. For the calculation of Tafel slope, overpotential was plotted against the log *j* [mA/cm$^2$]. This followed by lineare regression to find the linearity. Herein, we see dependence on the overpotential. In the low overpotential regime, Tafel slope for HER and OER is 122 and 73.74 mV/dec, respectively. However, in the mid-high overpotential range, the values of 243 and 291 mV/dec were observed for HER and OER, respectively. Song et al. [69] reported a similar trend in Tafel slopes. This anomaly from the simple linearity could be due to such factors as back reaction at low overpotentials, mass transport together with the blocking effect of produced $H_2$ bubbles at high overpotentials, formation of a large number of N–H moieties, and the dependence of adsorbed hydrogen intermediate on overpotential [70-72].

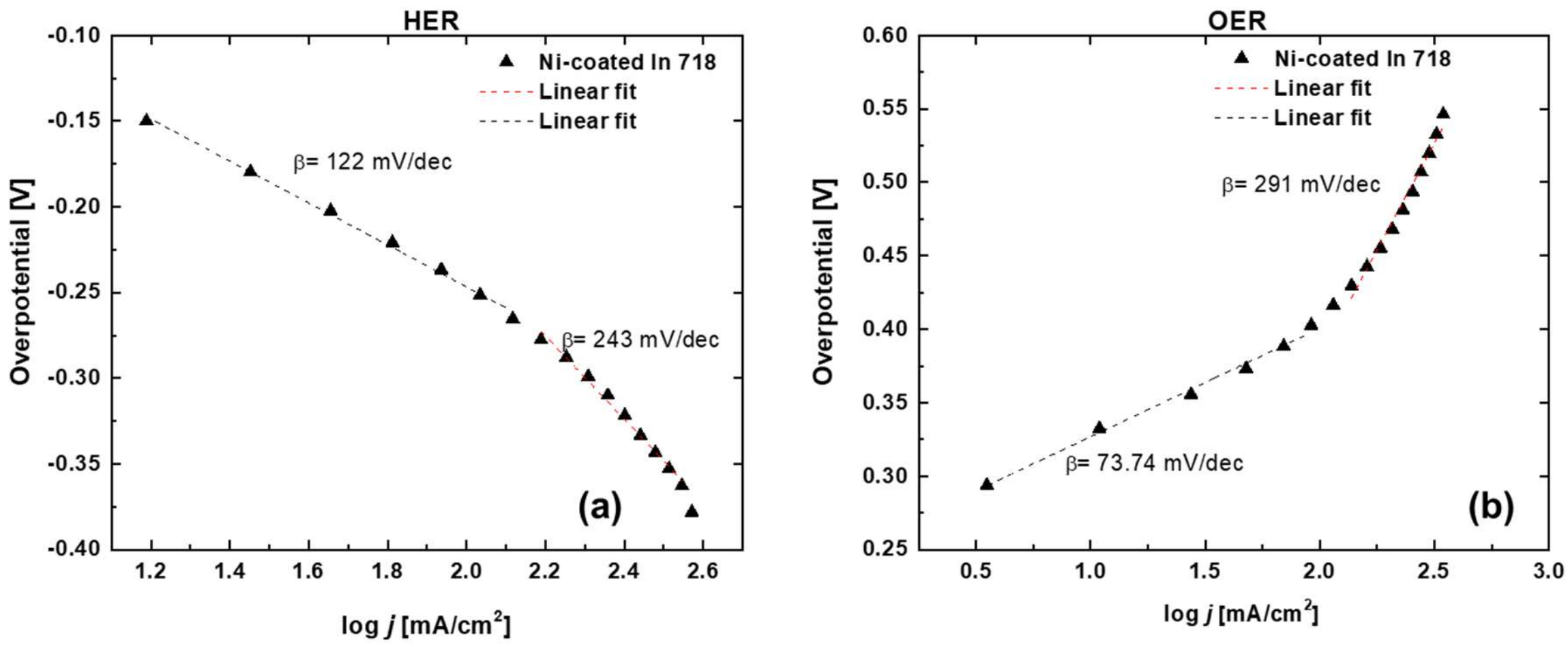

**Fig S8. Tafel slope for nickel-coated In718 (a) for HER; (b) for OER.**

## 8. Long- term test

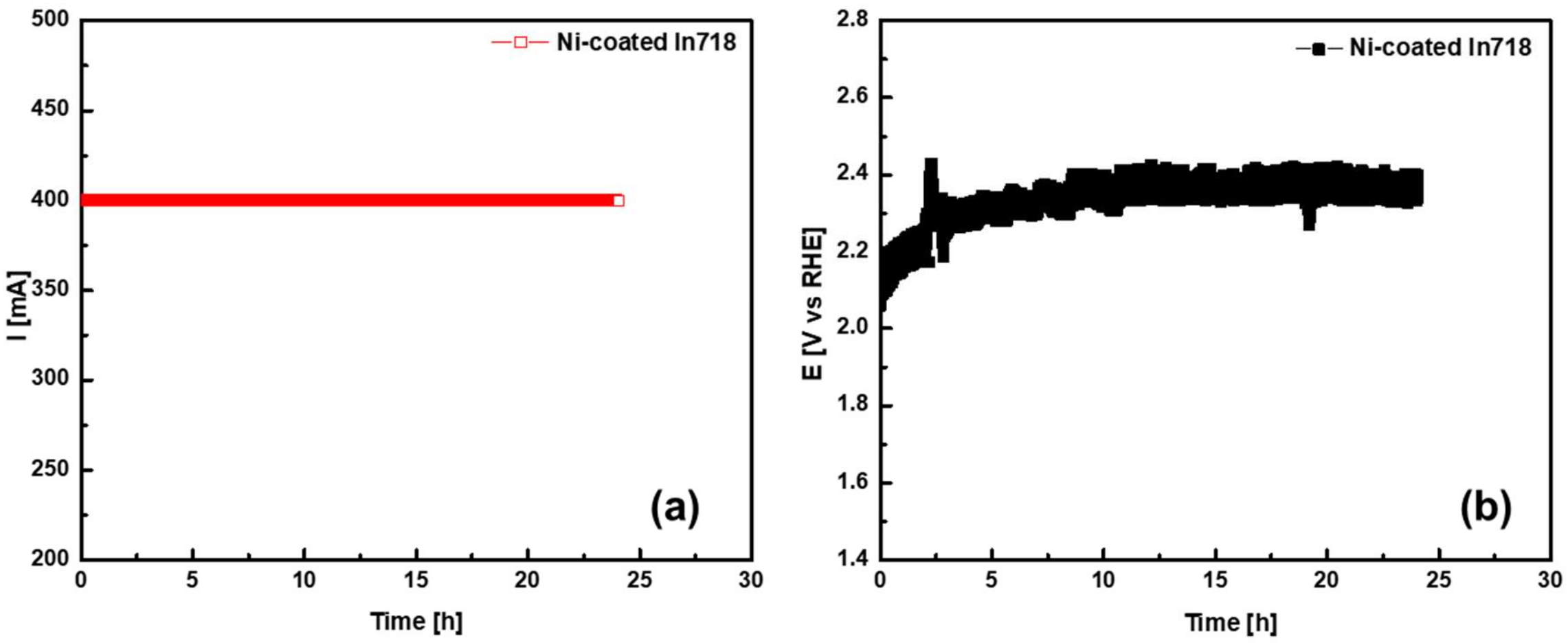

**Fig S9. Long term chropotentiometric experiment of 24h for OER (a) transient of constant applied current of 400mA; (b) corresponding transient of potential response with respect to time.**